\documentclass[acmlarge,dvipsnames]{acmart}

\usepackage{bm}
\usepackage{dsfont}
\usepackage{wrapfig}
\usepackage{xspace}
\usepackage{nccmath}
\usepackage{amsmath}
\usepackage{caption}
\usepackage{subcaption}
\usepackage{hyperref}
\usepackage{booktabs}
\usepackage{tabularx}

\AtBeginDocument{%
  }

\setcopyright{acmlicensed}
\acmJournal{IMWUT}
\acmYear{2025} \acmVolume{9} \acmNumber{3} \acmArticle{146}
\acmMonth{9}\acmDOI{10.1145/3749455}

\begin{document}

\title{Map as a By-product: Collective Landmark Mapping from IMU Data and User-provided Texts in Situated Tasks}


\author{Ryo Yonetani$^*$$^\dagger$}\thanks{$^*$ Corresponding author.}\thanks{$^\dagger$ Equal contribution.}
\affiliation{%
  \institution{CyberAgent}
  \city{Tokyo}
  \country{Japan}
  }
\email{yonetani\_ryo@cyberagent.co.jp}
\orcid{0000-0002-2724-6233}

\author{Kotaro Hara$^\dagger$}
\affiliation{%
  \institution{Singapore Management University}
  \city{Singapore}
  \country{Singapore}}
\email{kotarohara@smu.edu.sg}
\orcid{0000-0002-7893-6090}

\renewcommand{\shortauthors}{Yonetani and Hara}
\newcommand{\codename}{Collective Landmark Mapper\xspace}

\def\eg{{\it e.g.}}
\def\cf{{\it c.f.}}
\def\ie{{\it i.e.}}
\def\etal{{\it et al. }}
\def\etc{{\it etc}}
\renewcommand{\ss}[1]{^{(#1)}}
\newcommand{\RY}[1]{\textcolor{cyan}{(RY) #1}}
\newcommand{\KH}[1]{\textcolor{magenta}{(KH) #1}}

\begin{abstract}
This paper presents \codename, a novel \textit{map-as-a-by-product} system for generating semantic landmark maps of indoor environments. Consider users engaged in situated tasks that require them to navigate these environments and regularly take notes on their smartphones. \codename exploits the smartphone's IMU data and the user's free text input during these tasks to identify a set of landmarks encountered by the user. The identified landmarks are then aggregated across multiple users to generate a unified map representing the positions and semantic information of all landmarks. In developing the proposed system, we focused specifically on retail applications and conducted a formative interview with stakeholders to confirm their practical needs that motivate the map-as-a-byproduct approach. Our user study demonstrates the feasibility of the proposed system and its superior mapping performance in two different setups: creating a product availability map from restocking checklist tasks at a retail store and constructing a room usage map from office inspection tasks, further demonstrating the potential applicability to non-retail applications.
\end{abstract}


\received{1 May 2025}
\received[revised]{1 May 2025}
\received[accepted]{1 May 2025}

\begin{teaserfigure}
  \includegraphics[width=\linewidth]{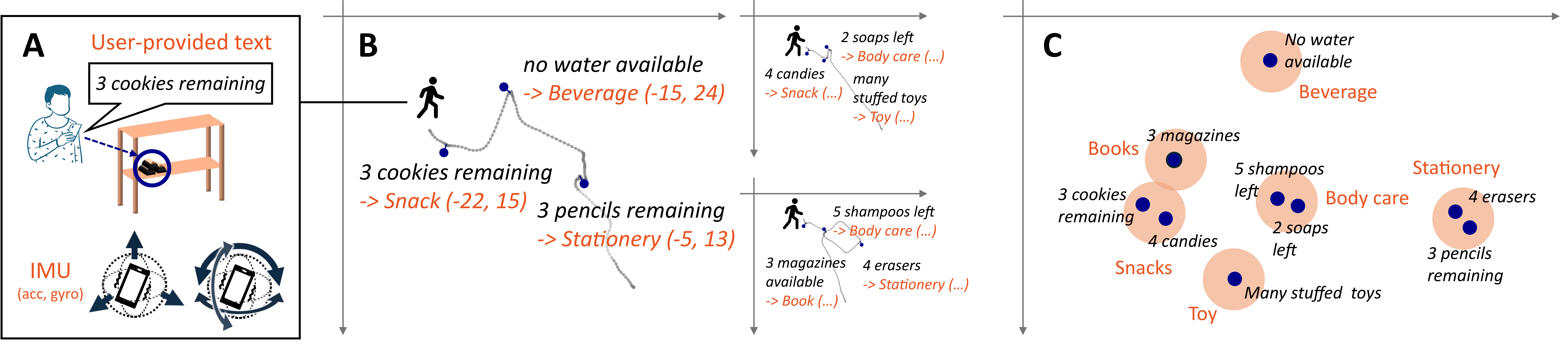}
  \caption{\textbf{Collective Landmark Mapping.} (A) We collect smartphone IMU data and free text inputs from users involved in indoor situated tasks. (B) Inertial navigation and large language models identify a small set of landmarks with their position and semantic information from each user's data. (C) A novel coordinate alignment optimization aggregates multiple user data to generate a unified semantic landmark map.}
  \Description{A figure with three panes. The first pane, marked as A, demonstrates a usage scenario, where a person is holding a smartphone and entering textual description about a landmark (shelf). This pane also illustrates two pictorial representations of sensors: accelerometer and gyro. 
  The second pane, B, shows three trajectories. Each trajectory represents how the user of the system walked in the indoor environment. Each trajectory is accompanied by annotations, representing what the user entered while maneuvering the space. For example, the first trajectory has three annotations, which says: ``3 cookies remaining -> Snack (-22, 15)'', ``no water available -> Beverage (-15, 24)'', and ``3 pencils remaining -> Stationery (-5, 13)''.
  The last pane, C, shows the map of aggregated landmarks. The figure shows six clusters, ``Beverage'', ``Books'', ``Snacks'', ``Toy'', ``Body care'', and ``Stationery''.
  }
  \label{fig:overview}
\end{teaserfigure}

\maketitle

\section{INTRODUCTION}
Location-tagged semantic data about indoor landmarks enriches location-based services for applications like in-store retail and office operations \cite{Melamed2016, Nurmi2011, andersson2020indoor, raychaudhuri2025semantic}.
When available, this data enables services to answer situational questions such as \textit{``Is there \underline{sufficient KitKat inventory} on the \underline{snack shelf}?''} and \textit{``Is \underline{meeting room 4.3} \underline{currently occupied}?''} (landmarks and semantic data underlined).
In addition to data about the presence of landmarks, creating location-based services requires their precise positions; some applications may even need richer semantic data that fluctuates dynamically.
Preparing such data is cumbersome and requires labor and financial costs that currently outweigh the benefits of indoor location-based services for user-facing applications and business operations \cite{deloitte2024future}.

\looseness=-1
We are particularly interested in location-based services for retail domains. Our interviews with retail store managers and part-time workers have revealed both challenges and opportunities to benefit from location-tagged semantic information (see Sec.~\ref{sec:interview} for more details). On the one hand, comprehensive knowledge about in-store product layouts, \ie, which product items are on which shelves, is critical for a variety of regular tasks such as customer assistance and inventory management. On the other hand, gaining such in-store knowledge takes a considerably long time (\eg, one month of training for newly-hired employees) despite regular changes in layouts due to new product introductions or sales promotions. We believe that location-based services would help retailers maintain up-to-date knowledge, enhancing customer experiences and reducing personnel training costs.

Much existing work has aimed to collect location-tagged semantic data by involving on-site crowd workers or volunteers, an approach often referred to as crowd or participatory sensing~\cite{Garg2019, Kim2018, Lin2016, Park2018, Zhou2021, li2022crowdsourcing, chon2012csp, Binalhaj2021}. However, in retail scenarios, it is not obvious whether retailers with low-profit margins can afford to recruit such workers or whether a limited number of employees, already burdened with a large volume of regular duties, can accommodate such additional tasks. Moreover, participatory sensing requires precise positional data to be associated with semantic data. Some approaches utilize off-the-shelf GPS~\cite{Kim2018, Lin2016}, which is suitable only for outdoor tasks. Indoor participatory sensing instead employs Simultaneous Localization and Mapping (SLAM)~\cite{li2022crowdsourcing}, WiFi localization~\cite{chon2012csp}, or inertial navigation combined with vision/wireless techniques~\cite{guo2014shopprofiler,teng2018sise,alzantot2012crowdinside}. These methods, however, necessitate extensive site surveys and regular maintenance of hardware devices, incurring additional financial and labor costs. The use of cameras further raises privacy concerns~\cite{akter2020uncomfortable} and increases energy consumption~\cite{naeem2020efficient}.

Motivated by retailers' potential needs and challenges for location-based services, we develop a prototype system, \codename, that exploits textual descriptions about landmarks generated during in-situ workers' regular tasks to create a \emph{map as a byproduct}.
For example, during retail inventory checks, workers patrol stores to identify low-stock items and create restocking checklists (\eg, \emph{``five KitKats remaining''}).
The key insight that enables our system is that such task-related text descriptions can serve as a source of semantic information about physical landmarks (\eg, shelves, rooms) when provided by workers.
As illustrated in Fig.~\ref{fig:overview}a, \codename utilizes a collection of IMU (inertial measurement unit; typically consisting of accelerometer and gyroscope) data and task-related free text inputs that are both available on the smartphones of users (\ie, local workers conducting their primary tasks).
Neural inertial navigation techniques~\cite{yan2020ronin} first estimate the user's movement trajectories from the IMU data.
Then, a state-of-the-art large language model (LLM) processes the user-provided texts to infer a semantic landmark label (\eg, `Snacks' category, `Meeting Room 3') associated with the nearby physical landmark. This label is further tagged with an estimated position derived from the trajectories as shown in Fig.~\ref{fig:overview}b.
The estimated trajectories do not perfectly align with the geometry of the indoor environment due to drift in inertial odometry and uncertainty in coordinate systems, resulting in initially inaccurate positions associated with these labels.
To address this, we create a novel optimization algorithm that refines the alignment of these positions by aggregating data across multiple users based on the similarity of the associated landmark labels. The output is a \emph{semantic landmark map} that contains refined positions representing the physical landmarks, tagged with their inferred labels and aggregated semantic information (\eg, product availability) provided by all users as shown in Fig.~\ref{fig:overview}c.

To validate the feasibility of our approach, we conducted a user study with 16 participants performing two indoor tasks in two separate environments: a restocking checklist task in a retail store and a room inspection task in an office space, with the latter demonstrating the potential applicability of our system to non-retail domains. 
The study revealed that \codename could provide accurate landmark maps with average positional errors of 0.8~m and 1.9~m for the respective environments, along with nearly perfect landmark coverage using data from 11 users. The results demonstrate that (i) the text data users generate through their primary tasks, when combined with IMU data, can aid in incidentally creating indoor landmark maps with semantic information such as product availability and room usage history, and (ii) our system can generate landmark maps with improved positioning accuracy and landmark coverage as the number of participating users increases. Furthermore, we conducted a simulation experiment to systematically evaluate the performance and limitations of our system under various factors, such as the number of landmarks, the variability of users' positions when visiting landmarks, and the presence of duplicated landmarks in the environment.

\section{BACKGROUND AND RELATED WORK}

Rich semantic information about the built environment makes location-based services integral to our lives.
Common location-based services include navigation, gaming, and other emerging applications such as advertising and safety (see \cite{huang2018location} for an extensive survey). 
They are more prevalent outdoors due to readily accessible localization technologies (\eg, GPS) and abundant semantic information about outdoor scenes (\eg, presence of businesses, traffic information) \cite{HAYWARD2022100608}.
While these applications would be valuable in indoor environments, the relative scarcity of up-to-date semantic information about indoor spaces limits their development. 
Therefore, this work contributes to the body of research on manual and (semi-)automatic methods for mapping semantic information about indoor environments.

\textcolor{black}{Building information and indoor landmarks are traditionally documented through manual processes and lack semantic data required to create location-based services \cite{Binalhaj2021, Ergen2021, Borrmann2018}. 
Modern companies involved in the construction and management of buildings increasingly use Building Information Modeling (BIM) to manage their physical layouts \cite{Borrmann2018}. 
Facility management may collect up-to-date data by asking occupants for feedback regarding comfort requirements and facility needs \cite{Ergen2021}. 
This data collection process largely relies on basic communication channels like phone calls and emails \cite{Ergen2021, Binalhaj2021}. 
While these methods serve basic facility management needs, they typically collect only basic facility features (\eg, room locations and elevator placements). 
Capturing rich and dynamic semantic information, such as inventory levels on shelves or office space utilization, requires substantial effort for frequent data collection and updates, making it a nontrivial task. There is a clear need for a streamlined approach to efficiently collect semantically detailed, application-relevant indoor spatial data.}

Research in automated indoor inspection is evolving beyond the traditional geometric mapping of indoor environments to semantic annotation of landmarks and spaces \cite{Rosinol2019, Cadena2016, Grinvald2019}. 
Visual Simultaneous Localization and Mapping (SLAM) enables a mobile robot to concurrently construct a geometric map of an unknown space and localize itself within that space \cite{Cadena2016}.
While conventional SLAM techniques focused on creating geometric maps, recent work by Rosinol \textit{et al.} \cite{Rosinol2019} and Grinvald \textit{et al.} \cite{Grinvald2019} integrates semantic labels captured by computer vision-based techniques with indoor geometric data. 
For example, Rosinol \textit{et al.} use pixel-level semantic annotation of images and automatically find correspondences with geometric data collected through SLAM \cite{Rosinol2019}. 
The semantic annotations of indoor environments collected through these methods may provide the data required for location-based services and applications mentioned above.
However, computer vision-based methods raise significant privacy concerns. 
Current technology indiscriminately collects all visible data, including sensitive or private moments of people who happen to be present (\eg, shoppers browsing items in a store). 
This lack of privacy-preserving computer vision approaches, combined with hardware and energy requirements, limits the widespread deployment of semantic SLAM technology.

In contrast to prior works, we seek an easy-to-deploy and low-cost approach by leveraging natural in-situ user activities to incidentally generate landmark maps as a by-product of existing data collection processes.
Our work shares motivation with existing mobile crowd sensing and participatory sensing approaches \cite{Garg2019, Kim2018, Lin2016, Park2018, Zhou2021, li2022crowdsourcing, chon2012csp}, particularly those that aim to minimize the data collection effort of individuals \cite{Kim2018}.
For example, in Kim \textit{et al.}'s on-the-go data collection method, workers receive task notifications in situations where they are likely able to contribute, minimizing changes to their normal behavior \cite{Kim2018}.
However, unlike outdoor scenarios benefiting from GPS \cite{Kim2018, Lin2016}, indoor participatory sensing faces localization challenges. Existing methods often employ WiFi positioning \cite{chon2012csp} that demands infrastructure deployment or Visual SLAM \cite{li2022crowdsourcing} with the deployment challenges mentioned above. Alternatively, inertial navigation systems~\cite{wu2019survey,hou2020pedestrian,woodman2007introduction}, particularly those using deep learning techniques~\cite{yan2020ronin,sun2021idol,liu2020tlio,herath2021fusion,herath2022niloc,saha2022tiny,zhou2023gps}, leverage ubiquitous smartphone sensors (\ie, IMU) to demonstrate robust localization capabilities independent of external infrastructure or visual data capture. Yet, the application of inertial navigation systems as the primary localization modality for indoor participatory sensing remains largely underexplored~\cite{guo2014shopprofiler,teng2018sise,alzantot2012crowdinside}.

Our contribution lies in developing an incidental crowd sensing method that combines inertial navigation, optimization, and user-provided text descriptions about indoor landmarks for positioning collected landmark data. Our approach does not rely on visual data that may limit deployment opportunities or environmental sensors that require regular maintenance and calibration.
Furthermore, our method is intended for use by individuals whose everyday tasks already include indoor environment inspection. It does not require additional conversations to be provided from users solely for localization unlike existing work~\cite{sheshadri2024conversational}, nor introduce any privacy violations beyond those inherent in users' existing routine responsibilities.

\section{FORMATIVE INTERVIEW}
\label{sec:interview}

To validate that incidentally collecting indoor map data would add value to indoor operations (\ie, what we call a \textit{map-as-a-by-product} approach), we conducted semi-structured interviews with three stakeholders: one store manager with seven years of experience across multiple retail stores, and two individuals with past experience as part-time workers in drug stores. A member of our research team grouped the emergent themes into related categories based on their natural relationships.

\subsection{Retailer's Needs}
\label{subsec:needs}
The interview results revealed several unmet retailer needs that could be addressed with indoor location data, as well as design constraints that data collection methods need to meet.

\paragraph{Item Localization and Identification}
\looseness=-1
Quickly finding the exact shelf location for specific product items is crucial for store maintenance tasks such as restocking missing items and checking expiration dates, as well as customer assistance like answering inquiries about product availability and guiding customers to products. Even knowing the general areas of items can be helpful for customer guidance. Identifying all locations where a single product might be stocked within the store would help avoid missing items during checks or replacements. While stakeholders acknowledged the benefits of collecting item location data, they shared concerns about the difficulty of identifying items just by visually scanning them (\eg, similar-looking packages, items stored in cardboard boxes). Furthermore, interviews highlighted that identifying which items need restocking often consumes significant time, requiring staff to visually scan shelves. A bottleneck also occurs in the backroom, described by interviewees as often chaotic and disorganized, where locating specific items (\eg, one particular brand of toilet paper among several similar ones) can be extremely time-consuming and relies heavily on individual staff knowledge, lacking standardization. Physical challenges, such as reaching high shelves for shorter staff, also add complexity to shelf-stocking tasks.

\paragraph{Staff Training}
Newly hired and temporary employees, including gig workers, must acquire knowledge of store layouts to perform essential retail functions such as customer assistance, merchandise reorganization, and inventory management. 
This requirement extends to experienced, regular staff when retail environments undergo frequent spatial reconfiguration.
Our interviews revealed that training personnel to achieve sufficient familiarity with store organization involves a time-intensive process requiring approximately one month of training. 
One key stakeholder expressed a desire for a system to facilitate location-based knowledge acquisition for staff members.
Such a system would serve dual purposes: (i) training new employees about product locations and (ii) supporting experienced staff in identifying unfamiliar positions of newly introduced merchandise. 
Stakeholders anticipated that implementing this technology could reduce training duration and minimize errors (\eg, inaccurate communication regarding product availability or location to customers).

\paragraph{Managing Layout Changes}
Retail environments undergo frequent layout modifications, including semi-annual major reorganizations, quarterly seasonal adjustments, and minor updates such as new product introductions, discontinued items, and price changes. 
These modifications require staff to maintain comprehensive knowledge of item locations to facilitate the movement of products between shelves. While major planograms might be set by headquarters, managers often have discretion over end caps and promotional displays, leading to more frequent, smaller-scale changes. The interviews also described how unsold or replaced items are often placed unsystematically in backroom storage (\eg, an `unsold items box'), making later retrieval difficult if needed.
However, as discussed above, training personnel on store layouts consumes significant time.
Moreover, even well-trained staff frequently possess imperfect knowledge of precise item locations. 
Therefore, a tool that provides on-demand information about item whereabouts would effectively support frequent layout updates for both experienced and less-experienced store staff.

\paragraph{Operational Efficiency}
Stakeholders expressed that comprehensive locational data of retail merchandise could serve as a foundation for future technologies addressing workforce shortages.
The retail sector currently faces personnel shortages, and thus, the industry demands efficiency-optimizing technologies that facilitate task completion with reduced staffing levels. The significance of restocking and shelf-facing tasks was underscored, with one interviewee estimating it constitutes up to 90\% of their non-cashier duties. Time lost due to inefficiencies stemming from handling large customer purchases (\eg, stockouts requiring apologies, staff tied up assisting with loading), and other demanding tasks like periodic inventory checks and complex customer interactions (\eg, handling complaints or language barriers) further strain limited staff resources and highlight the need for efficiency gains.
The amount of time employees can stay in the store outside of business hours is limited (\eg, less than an hour), leaving little room for them to work beyond their normal duties such as cash register closing, store cleaning, or restocking checks.
For instance, they noted that tools supporting logistical optimization of merchandise replenishment pathways based on up-to-the-moment information on item stock levels would improve operational efficiency.
This theme not only identified the desire for tools to make day-to-day retail operations more efficient but also highlighted design constraints for technologies that collect locational data of products.
The interviewees emphasized that data collection tools should not introduce additional operational processes but should be integrated into existing essential functions.
They also noted a preference for smartphone-based applications over tablet interfaces (too large for practical day-to-day use) or specialized smart cart systems.

\subsection{Design Goals}
From the above summary of the interview results, we identify the following design goals for the Collective Landmark Mapper system:

\begin{itemize}
    \item \textbf{Data Requirements.} 
    The data collected by the system should support training novice staff about item layouts and enhance day-to-day retail operations. 
    It should enable the development of tools that reduce the learning curve for new and temporary employees while facilitating efficient item localization for stocking, customer service, and internal monitoring procedures such as expiration date verification. 
    Although static floorplans might be available to describe the overall shelf arrangement, retail operations necessitate more fine-grained and dynamic information about product availability at each shelf.
    Consequently, precise and up-to-date item locations at the shelf-by-shelf level are essential.
    
    \item \textbf{Leverages Existing Workflow and Tools.} 
    In designing the system, we must acknowledge the current labor constraints in retail environments. 
    Solutions that demand additional dedicated staff time for comprehensive item mapping are also impractical.
    Therefore, the system should integrate seamlessly with existing operational workflows.
    
    \item \textbf{Easy to Deploy.} Through our interviews, we confirmed that the challenges discovered occur across multiple retail companies and stores. These challenges require more than a temporary upgrade to a single storefront to be effectively addressed.
    \textcolor{black}{Introducing in-store apps or dedicated systems to manage and display product locations, while already available in some major retail chains, is not necessarily acceptable as it will incur substantial installation costs and inevitably impact overall store management.}
    Compared to alternatives requiring cameras or other hardware (WiFi/SLAM surveys), the reliance on smartphone IMU and text input is easier to deploy.
\end{itemize}

The map-as-a-by-product paradigm embodies an efficient solution addressing the above design goals, wherein staff members collect data during routine restocking and inventory verification processes, eliminating the need for separate data collection procedures. 
Leveraging existing smartphone devices, possibly those already owned by workers, minimizes overall deployment and maintenance costs.

\begin{wrapfigure}[15]{r}{0.25\linewidth}
\centering
\vspace{-3em}
\includegraphics[width=\linewidth]{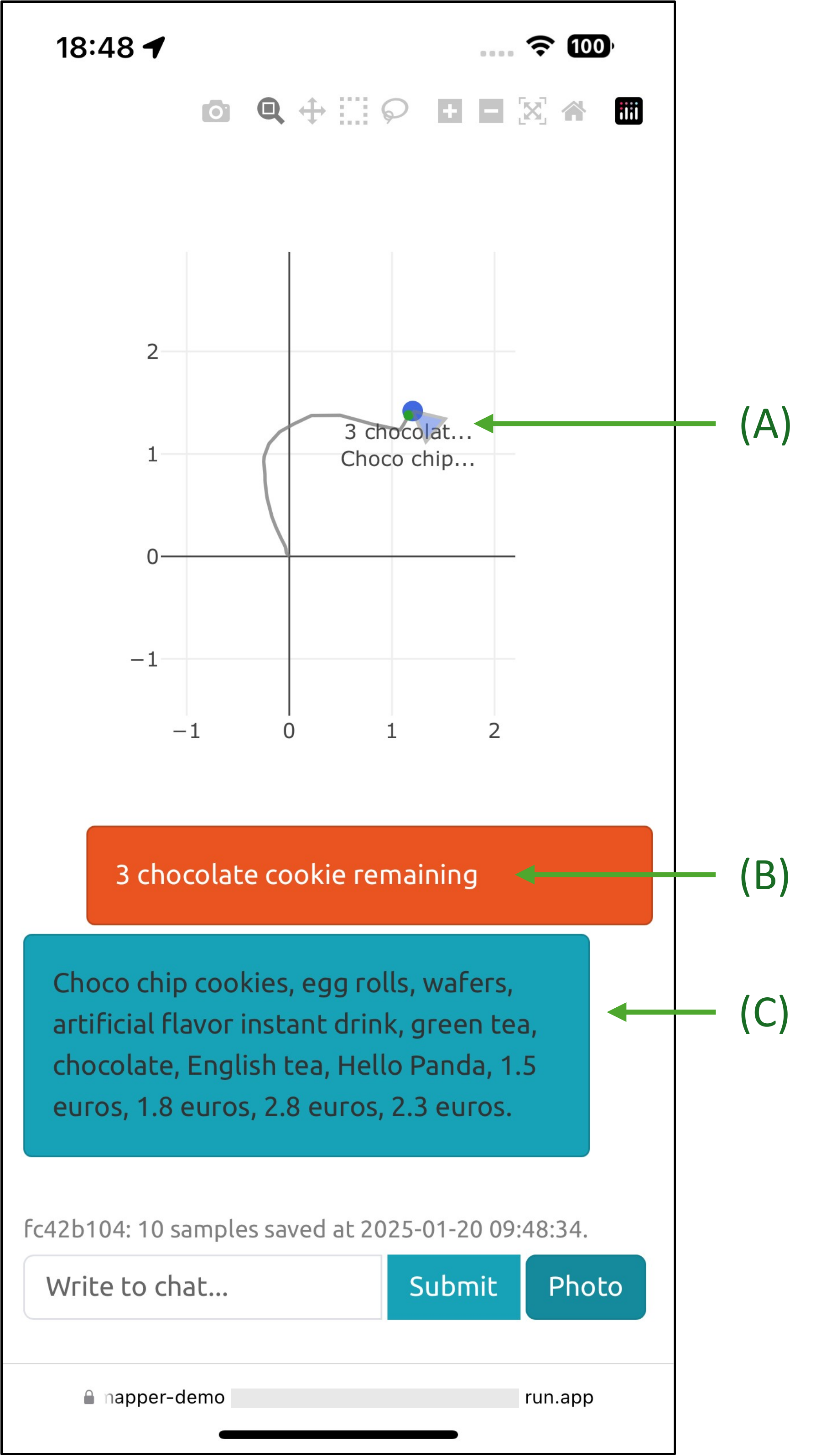}
\caption{Data collection app. (A) Trajectory visualization and annotations. (B \& C) User-provided and LLM-generated landmark descriptions.}
\Description{On the trajectory visualization, there is one path representing the path that the user walked, with a blue dot at the end showing the user's current position. Along the path, there is one annotation, saying ``3 chocolate... Choco chip...'' Below the visualization, there is an orange chat bubble that says ``3 chocolate cookie remaining''. This represents the user's manual input. Below that, there is a blue bubble. It says, ``Choco chip cookies, egg rolls, wafers, artificial flavor instant drink, green tea, chocolate, English tea, Hello Panda, 1.5 euros, 1.8 euros, 2.8 euros, 2.3 euros.'' This represents a description generated by the LLM.}
\label{fig:ui}
\end{wrapfigure}

\section{COLLECTIVE LANDMARK MAPPER SYSTEM}
Based on the design goals derived above, we have created \codename, which consists of three parts: (i) a mobile application that gathers and visualizes user movement and free text inputs during situated tasks, (ii) a back-end module for identifying landmarks from user-provided data, and (iii) a module that uses an optimization algorithm to align the collected landmark data across multiple users. The generated \emph{semantic landmark map} is a collection of location-tagged spots with semantic information, such as \emph{``snack shelf (x=4.8 m, y=5.6 m, available products: 3 cookies, 4 candies, ...)''} (see also Fig.~\ref{fig:overview}c).

\subsection{Data Collection App}
\label{sec:data_collection_app}
Our data collection app implements the following features:

\begin{itemize}
    \item \textbf{Inertial Movement Tracking (Fig.~\ref{fig:ui}A)}: The app tracks IMU data (\ie, acceleration and angular velocity) and visualizes the movement data at 2 fps. The visualization shows the user trajectory inferred from the IMU data using neural inertial navigation~\cite{yan2020ronin}, the heading direction obtained with quaternion-based orientation filtering~\cite{laidig2023vqf}, and text annotations.
    \item \textbf{Text Input (Fig.~\ref{fig:ui}B)}: The app provides a text input field that accepts free text from the user. We expect users to provide descriptions of in-store landmarks' states, such as \textit{``3 chocolate cookies remaining''} or \textit{``Room 3 is occupied.''} The UI lists the provided text inputs as a chat history and as textual annotations.
    \item \textbf{Photo Capture and Analysis (Fig.~\ref{fig:ui}C)}: We also implemented a visual recognition feature that allows users to take photos and send them to an external LLM service\footnote{GPT-4o mini on Microsoft Azure OpenAI Service (\url{https://learn.microsoft.com/en-us/azure/ai-services/openai/concepts/models}), selected for its balance of capability and cost-effectiveness.} to describe the visual scene. The descriptions are presented in the chat history and as annotations. \textcolor{black}{Note that this photo-capturing feature is not necessary for our data collection approach, which only collects IMU and text data without privacy concerns. We implemented it to compare our system against vision-based baselines, which we detail further in Sec.~\ref{sec:alg_implementation}.}
\end{itemize}

\subsection{Landmark Identification}
\label{sec:landmark_identification}
Our landmark identification module identifies the names of landmarks, such as the categories of products on each shelf, from user-provided text inputs, as well as their positions obtained from movement tracking results. We use an LLM with a custom prompt to infer landmark names. For example, for the restocking checklist tasks conducted in our user study, we used the following prompt: \textit{``Given the list of shopper's descriptions of items and their availability in a grocery store, describe the corresponding list of item categories (one of the following: `Stationery', `Beverages', `Snacks', ...). Description: 3 chocolate cookies remaining.''} Here, we expect the module to extract the keyword `chocolate cookies' and infer that the category is `Snacks'. The position associated with this inferred label corresponds to the user's estimated location while stationary at the landmark, typically averaged over the duration of the stop detected by the inertial navigation system. We then use the identified landmark names and positions in the subsequent coordinate alignment optimization. Currently, the system assumes reasonably clear text input; robust handling of highly ambiguous or erroneous user descriptions is considered future work.

\begin{wrapfigure}{r}{0.25\linewidth}
\centering
\vspace{-2em}
\includegraphics[width=\linewidth]{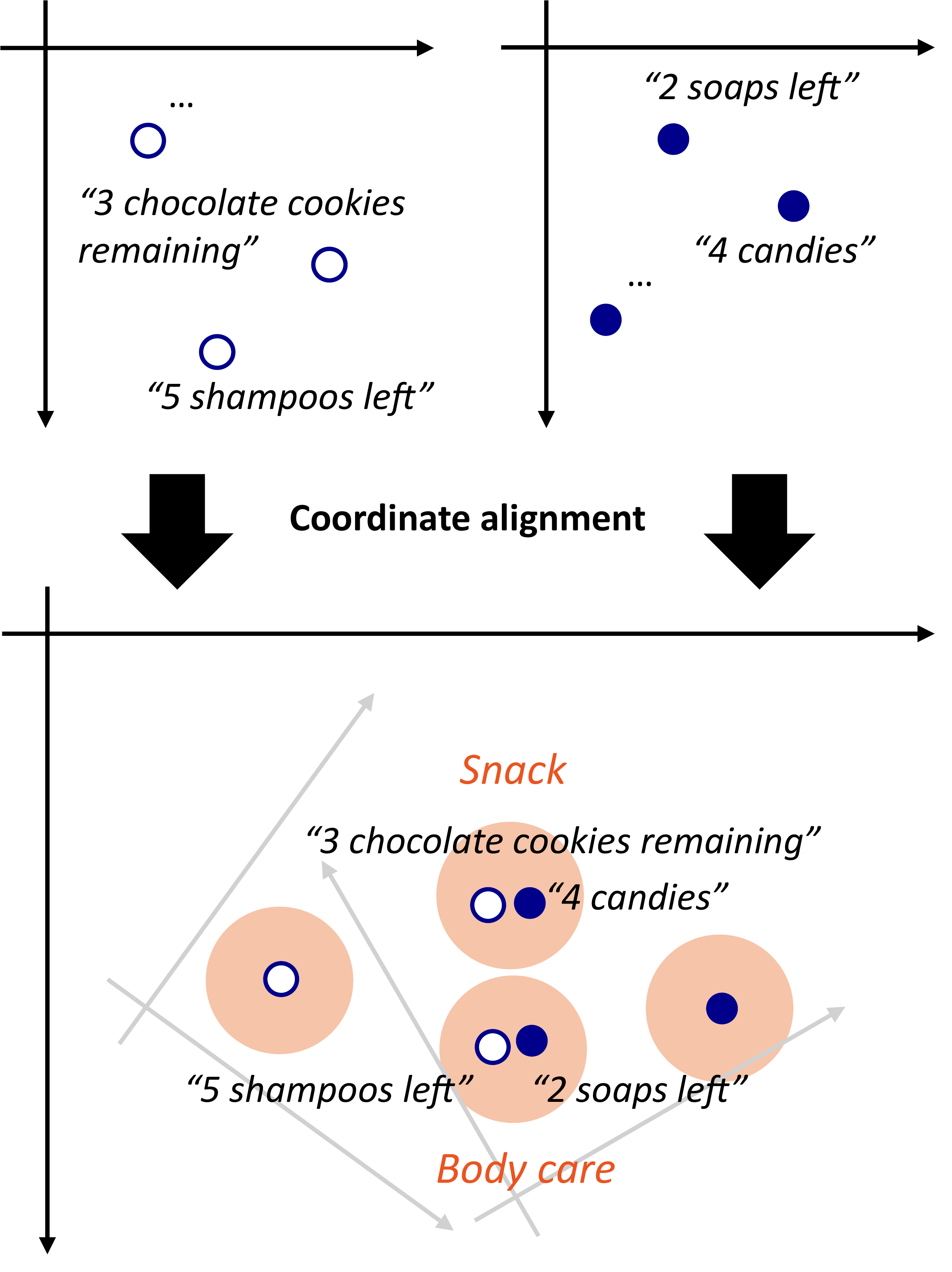}
\caption{Coordinate alignment optimization. Multiple recordings with different starting locations and orientations are transformed to be aligned onto a shared coordinate system.}
\Description{There are two rows of images. On the first row, there are two sets of landmarks, each of them mapped on its own Cartesian plane. The bottom row shows one Cartesian plane with the landmarks from the two sets of the landmarks above combined into one set and mapped together.}
\label{fig:optimization}
\end{wrapfigure}

\subsection{Coordinate Alignment Optimization}
\label{sec:coordinate_alignment}

Our key technical contribution is a coordinate alignment module that aggregates landmark data provided by multiple users to generate a unified semantic landmark map. Importantly, the coordinate systems used to represent landmark positions are unknown and different for each recording, depending on the starting location and orientation of a smartphone in the environment. To aggregate multiple recordings, we must align them onto a shared coordinate system.

To this end, we developed a novel optimization algorithm for coordinate alignment. As illustrated in Fig.~\ref{fig:optimization}, our algorithm computes a transformation (rotation and translation) for each recording to ensure `related' landmarks (\eg, products of the same category) are positioned closer together in the unified map. Once optimized, related landmarks are aggregated on the shared coordinate system to derive their median position.

Formally, this optimization problem is defined as the minimization of the objective function:
\begin{fleqn}[60pt]
\begin{align*}
\mathcal{L}=\sum_{i\in \mathcal{N}} \sum_{j\in \mathcal{N}} S_{i, j} \left\|\bm{q}_i - \bm{q}_j\right\|^2_2.
\label{eq:objective}
\end{align*}
\end{fleqn}
Here, $i, j \in \mathcal{N}$ ($i \neq j$) represent the indices of landmarks, and $\mathcal{N}$ is the index set of all landmarks gathered from participating users. $S_{i, j} \in [0, 1]$ is a relatedness score between the $i$-th and $j$-th landmark names introduced below. $\bm{q}_i$ and $\bm{q}_j$ are the transformed landmark positions on the shared coordinate system, given by $\bm{q}_i = R_{\Omega_i}\bm{p}_i + \bm{t}_{\Omega_i}$, where $\bm{p}_i$ is the original position in the $\Omega_i$-th recording (with its unique coordinate system), and $R_{\Omega_i}$ and $\bm{t}_{\Omega_i}$ are the rotation matrix and translation vector to be optimized. We minimize this objective function $\mathcal{L}$ with respect to $\{R_{\Omega_i},\bm{t}_{\Omega_i}\mid i\in [1,\dots,N]\}$ via gradient descent until convergence to obtain all the landmark positions $\{q_i\mid i\in[1,\dots,N]\}$.

\subsection{Algorithm Implementation}
\label{sec:alg_implementation}

In calculating the relatedness score, $S_{i, j}$, we compared our approach against two vision-based baselines:
\begin{itemize}
    \item \textbf{User-provided texts.} In our proposed approach, the relatedness score between identified landmark names is defined by first using a text embedding model\footnote{text-embedding-3-small on Microsoft Azure OpenAI Service (\url{https://learn.microsoft.com/en-us/azure/ai-services/openai/concepts/models}), chosen as a standard and cost-effective text embedding model.} to convert the names into vectors and then calculating their distance.
    \item \textbf{Image captioning.} We implemented a baseline method leveraging state-of-the-art vision-based semantic analysis, which infers landmark names from descriptions of items and texts automatically recognized in photos (as introduced in Sec.~\ref{sec:data_collection_app}) rather than from user-provided texts.
    \item \textbf{Image feature matching.} As another baseline, we investigated whether conventional image feature matching could be used to measure relatedness. While adopting the image captioning approach above to infer landmark names, we calculated relatedness scores using feature vectors extracted from photos with a pre-trained ResNet~\cite{he2016deep}, a standard choice for image feature extraction.
\end{itemize}

\section{USER STUDY}

\subsection{Participants and Tasks}
\textcolor{black}{To evaluate the proposed system in a realistic setting, we conducted our user study in real retail and office environments during normal business hours.}\footnote{\textcolor{black}{Although the proposed approach is designed to collect data during regular operations, our system developed in this study is a prototype for feasibility studies and is not currently integrated into applications used in actual operations.}} We recruited 16 participants (13 male and 3 female) through personal contacts and word of mouth. These participants were selected based on their ability to walk without difficulty and their familiarity with smartphone use and the environments where the tasks were conducted. Each participant was assigned one of the following two tasks:
\begin{itemize}
\item \textbf{Restocking checklist.} We asked eight participants to walk through a retail store (approximately 70 m$^2$ with 28 shelving units, \textcolor{black}{which is a typical size for small grocery stores commonly found in Asian cities, assuming a standard shelving unit size of 1.2~m $\times$ 0.9~m and an aisle width of 1.8~m}) to check product availability. They were asked to visit three to five randomly picked shelves out of ten shelves selected for evaluation (\eg, oral care, snacks, cosmetic goods). At each shelf, they selected a random product, freely described its remaining stock (\eg, \textit{``OCH Tune toothbrush, only 1 left''}). To enable comparison with vision-based baseline methods, participants also took a single photo of the shelf using the app. By aggregating these data, we constructed a product availability map as a semantic landmark map, which describes the positions of shelves and the recent status of product availability noted by the participants.
\item \textbf{Office inspection.} We asked the other eight participants to navigate an office space (approximately 300 m$^2$ with 30 rooms) to investigate whether reserved rooms were actually in use. They were randomly assigned to investigate three to six out of 14 rooms. Upon arriving in front of a room, they used our data collection app to freely describe if anyone was in the room (\eg, \textit{``Room 3 is occupied''}). To enable comparison with vision-based baseline methods, participants also took a photo of the room number on the door. With this task, we aimed to show that the proposed system can potentially be extended to non-retail domains such as office operation improvement. Similar to the product availability map introduced above, we constructed a room usage map that records the position and recent usage history of the rooms.
\end{itemize}
\textcolor{black}{Note that we limited the number of shelves/rooms to visit by participants to minimize interference with ongoing business operations. The effect of varying the number of visited landmarks is investigated through a simulation study detailed in Sec.~\ref{sec:simulation}.}

After being briefed on the task and how to use the data collection app, participants repeated their assigned tasks five times, while changing the rooms/shelves they visited each time. We instructed participants not to point their phone cameras at people around them when taking photos for vision-based baselines, and not to include people in their photos. Each recording lasted approximately 3-5 minutes, \textcolor{black}{and the total experiment time for each participant was less than 30 minutes. This is within realistic workloads particularly for retail scenarios, as we confirmed in interviews with retailers that they have around 1 hour outside business hours to complete various store closing operations including restocking checks.} After removing invalid data due to unstable data communication or API failure, we obtained 39 and 36 valid records for the restocking checklist and office inspection tasks, respectively. 

\subsection{Evaluation Metrics}
We used positional error and coverage as evaluation metrics to assess the accuracy of landmark mapping using our method and algorithm with the three relatedness metrics. For each landmark, we calculated positional error based on the Euclidean distance between its mapped and ground truth positions. We obtained the ground truth landmark positions by manually annotating the center of each target shelf (for the retail store) or room door (for the office space) on the provided building floor plans. We aligned the generated landmark map coordinates to the ground truth coordinates using an optimal similarity transformation (allowing rotation, translation, and uniform scaling) computed via Singular Value Decomposition (SVD) to minimize the least-squares error between corresponding landmark pairs. The positional error was then calculated as the average Euclidean distance between the aligned generated landmark positions and their ground truth counterparts. To evaluate performance, we averaged the scores across ten random trials using different record sets, testing aggregation levels from 1 record (no aggregation) to 11 records. We measured coverage by counting the number of landmarks identified. Additionally, we recorded the time from visiting a landmark to the completion of text entry (\ie, regarded as part of regular tasks), as well as the time to take a photo and confirm the LLM's captioning result to evaluate the additional time required by vision-based mapping baselines.

\subsection{Result}
\begin{figure}[t]
\centering
\begin{subfigure}[b]{\linewidth}
\includegraphics[width=\linewidth]{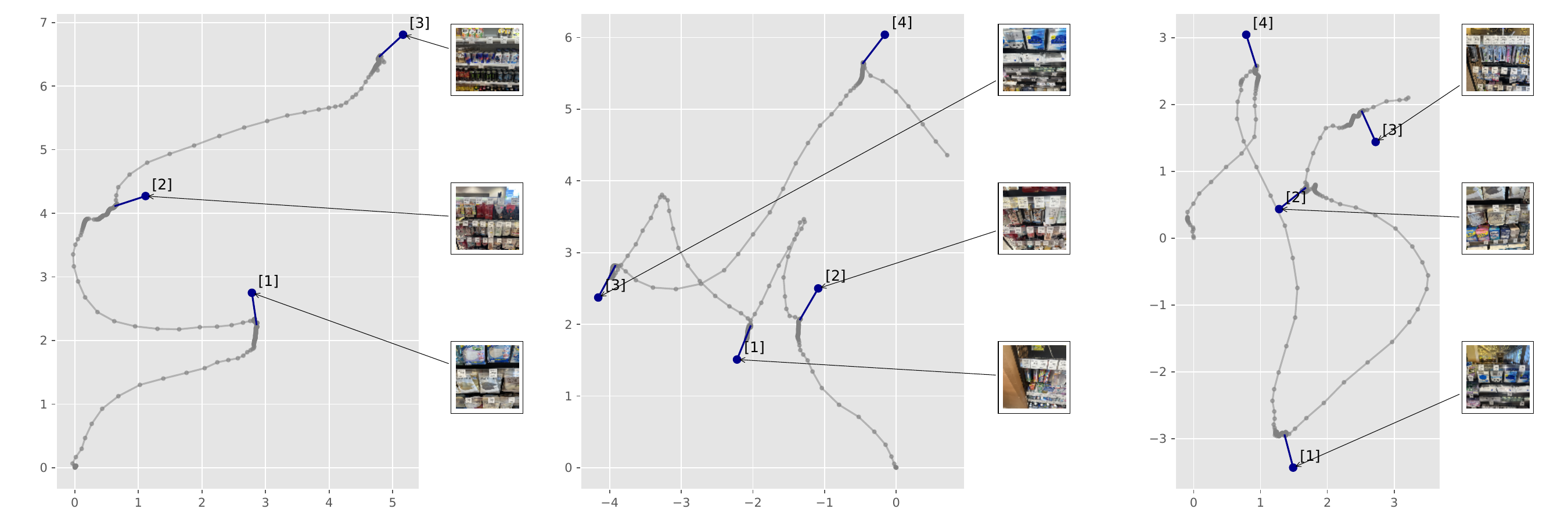}
\end{subfigure}
\begin{subtable}[h]{\linewidth}
\centering
\scriptsize
    \begin{tabularx}{\linewidth}{XXX}
[1] \textcolor{OrangeRed}{(User-provided) Ultra-comfortable smart color mask - 1 left.}
\textcolor{RoyalBlue}{(Image caption) Standard size masks (7 pieces), smart color masks, 3D masks, 398 yen, 468 yen, 7 pieces each.}
[2] \textcolor{OrangeRed}{(User-provided) My lash mascara remaining: 3.}
\textcolor{RoyalBlue}{(Image caption) Transino Medicinal Whitening Essence, 1500 yen, Transino Medicinal Whitening Facial Mask, 1900 yen, ONE BY KOSE Melano Shot White D, 950 yen, Wonder Honey Blissful Essence Gel Oil, 1100 yen, ONE BY KOSE Melano Shot White D, 950 yen, Bright Age Lift White UV Base CC, 1800 yen, Esprique Comfort Make Cream, 900 yen, Risage Beaute Medicinal Wrinkle Repair Lotion, 1300 yen, Ettusais Face Edition Primer, 1200 yen, Rohto Skin Lab Extreme Moist Hyaluronic Liquid, 1300 yen, Elixir Superieur Lift Moist Emulsion, 2000 yen.}
[3] \textcolor{OrangeRed}{(User-provided) Plenty of Red Bull left.}
\textcolor{RoyalBlue}{(Image caption) A day's worth of vegetables, gentle barley tea, BOSS, Red Bull, Monster, EX, price tag 108 yen, price tag 118 yen, price tag 159 yen, price tag 198 yen, price tag 229 yen.}
&
[1] \textcolor{OrangeRed}{(User-provided) OCH Tune toothbrush - 1 left}
\textcolor{RoyalBlue}{(Image caption) Toothbrush, gum dental brush, Clear Clean toothbrush, Clinica Advantage toothbrush, drugstore price tags: 298 yen, 328 yen, 338 yen, 398 yen.}
[2] \textcolor{OrangeRed}{(User-provided) W eyebrow gel - 1 left.}
\textcolor{RoyalBlue}{(Image caption) Meishoku eye cream, natural- based, moisturizing bath additive, Meishoku beauty soap, refreshing gel, medicated whitening cream, Clear Clean Premium, pure white, moisturizing lotion, all-in-one beauty serum, face cream.}
[3] \textcolor{OrangeRed}{(User-provided) Silver Ion Wet Towel Tissue, 6 pieces}
\textcolor{RoyalBlue}{(Image caption) 99\% Water Wipes, 108 yen, 99\% Water Wipes, Elleair + Water, You can feel the difference with one wipe, No. 1, 4-pack, Toilet Paper.}
[4] \textcolor{OrangeRed}{(User-provided) Turmeric Power - Turmeric Extract, 3 pieces}
\textcolor{RoyalBlue}{(Image caption) Monster Energy, 213 yen, Real Gold, Oronamin C, Thiovita Drink, Lipovitan D, Ucon no Chikara, ZONe, Thiovita Gold, Fight Ippatsu, Thiovita Sparkling, Lipovitan Fine, Thiovita Drink Eye, Vitamin Drink 30 yen discount, Gold, E-Scup.}
&
[1] \textcolor{OrangeRed}{(User-provided) Nose Celeb tissues remaining: 1}
\textcolor{RoyalBlue}{(Image caption) Giraffe Sunset car wet cloth, squeaky dishcloth, Japan-made antiviral mask, Iris Ohyama disposable mask, Moss Protection mask, medical surgical mask, 164 yen, 218 yen, 148 yen, 448 yen, 328 yen, 598 yen, 578 yen, 498 yen.}
[2] \textcolor{OrangeRed}{(User-provided) Masmix mask remaining: 5}
\textcolor{RoyalBlue}{(Image caption) MAS MIX, FACE FIT, be-style, Fitty, Ultra Comfort, Sarafit, Ultra 3D, RUKUKA MASK}
[3] \textcolor{OrangeRed}{(User-provided) Mitsubishi Jet Stream 05 Red Ballpoint Pen, 1 remaining.}
\textcolor{RoyalBlue}{(Image caption) Utility knife with replaceable blade, smart key holder, pen- shaped mini scissors, stapler, masking tape cutter, tape roller, slim mini stapler, card case, compact stapler, slim smart case.}
[4] \textcolor{OrangeRed}{(User-provided) Zone energy drink, 11 remaining.}
\textcolor{RoyalBlue}{(Image caption) Soy milk, banana-flavored soy milk, Marusan, Green Zone, Perfect Vitamin, yuzu, ZONe, ZONe INFINITE BLACK, energy drink, Monster Energy, Monster, bottle, can, 65 yen, 158 yen, 159 yen, 2 for 258 yen, 50\% bonus, price, quantity, 1 can, 500ml, 250ml, ZONe MAD HACK, energy drink, 450ml, 130 yen, 500 yen, 100 yen, 10\%, 198 yen, bonus, tax- inclusive price, juice, product, Perfect Vitamin, yuzu jelly.}
\end{tabularx}
\end{subtable}
\caption{Examples of collected data in the retail restocking checklist tasks: movement trajectories (gray lines), positions of identified landmarks (blue points), some taken photos (machine-translated), user-provided texts (red texts), and image captions generated from the photos (blue texts). Each trajectory has its own origin $(0, 0)$ and potentially varying true-north; this requires our alignment algorithm for creating a unified semantic landmark map.}
\Description{The figure consists of three trajectory plots, each depicting a path taken in a spatial environment. The paths are marked with gray lines representing the movement and blue dots indicating specific points of interest along the trajectory. Each point is labeled numerically and connected to an image that visually represents the scene or object at that location. In the first plot, three points are labeled [1], [2], and [3], each associated with an image showing shelves or products. The trajectory follows a curved path with these points positioned sequentially along the path. The second plot features four labeled points [1], [2], [3], and [4], each connected to images of shelves, similar to the first plot. The trajectory in this plot is more complex, with multiple curves and intersections. The third plot also includes four labeled points [1], [2], [3], and [4], linked to images of shelving or products. The path in this plot demonstrates a winding trajectory with several loops and turns.}
\label{fig:store_inputs}
\end{figure}

\begin{figure}[t]
\centering
\begin{subfigure}[b]{\linewidth}
\includegraphics[width=\linewidth]{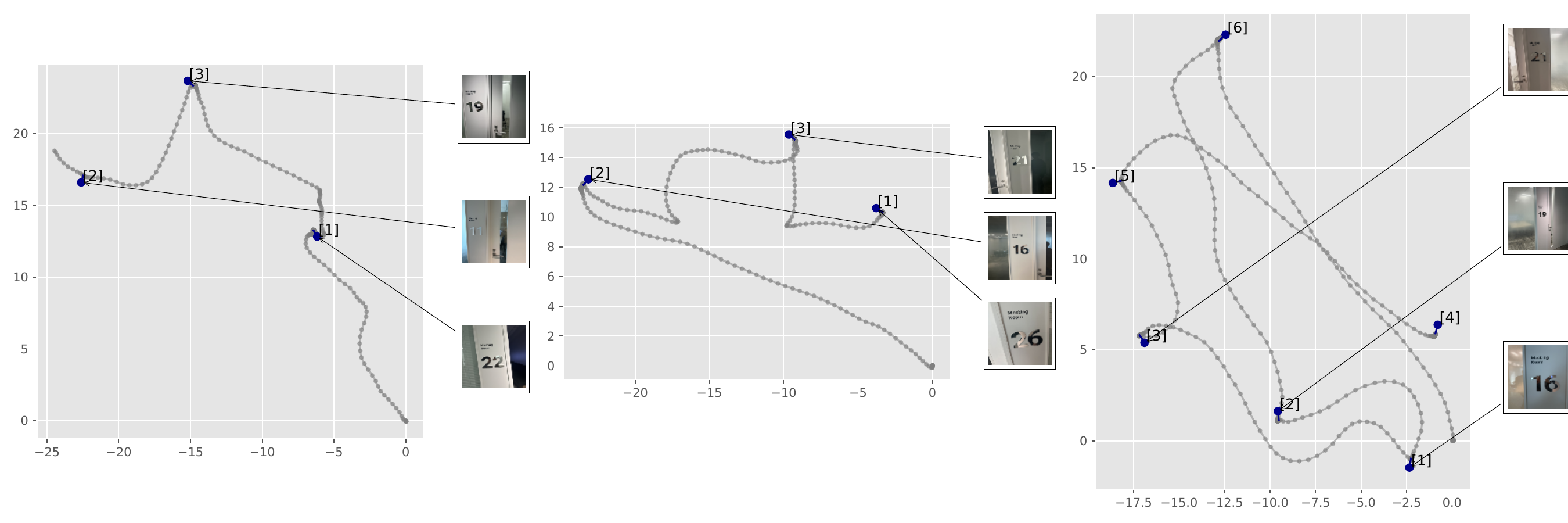}
\end{subfigure}
\begin{subtable}[h]{\linewidth}
\centering
\scriptsize
    \begin{tabularx}{\linewidth}{XXX}
[1] \textcolor{OrangeRed}{(User-provided) Room 22 available}
\textcolor{RoyalBlue}{(Image caption) Meeting Room 22, Door}
[2] \textcolor{OrangeRed}{(User-provided) Room 11 available.}
\textcolor{RoyalBlue}{(Image caption) Meeting room, 11, door, door handle, chair.}
[3] \textcolor{OrangeRed}{(User-provided) Room 19 available}
\textcolor{RoyalBlue}{(Image caption) Meeting room door, doorknob, chair, text "Meeting Room," number "19."}
&
[1] \textcolor{OrangeRed}{(User-provided) 26 vacancies}
\textcolor{RoyalBlue}{(Image caption) "Meeting Room 26"}
[2] \textcolor{OrangeRed}{(User-provided) 16 unused}
\textcolor{RoyalBlue}{(Image caption) Meeting room sign, number 16, door, door handle, conference room.}
[3] \textcolor{OrangeRed}{(User-provided) Room 21 available}
\textcolor{RoyalBlue}{(Image caption) Meeting Room 21, door handle, glass door, silhouette.}
&
[1] \textcolor{OrangeRed}{(User-provided) Utilization of 16}
\textcolor{RoyalBlue}{(Image caption) Meeting room, 16, door, silhouette.}
[2] \textcolor{OrangeRed}{(User-provided) No use of 19}
\textcolor{RoyalBlue}{(Image caption) Meeting room, 19, door, table, chair, ceiling light, window glass, carpet floor.}
[3] \textcolor{OrangeRed}{(User-provided) No use of 21}
\textcolor{RoyalBlue}{(Image caption) Door, door handle, number 21, text "Meeting Room," whiteboard, glass door.}
[4] \textcolor{OrangeRed}{(User-provided) No use of 9.}
\textcolor{RoyalBlue}{(Image caption) Meeting Room 09, doorknob}
[5] \textcolor{OrangeRed}{(User-provided) No usage of 23.}
\textcolor{RoyalBlue}{(Image caption) Emergency exit, door, doorknob, glass wall, Meeting Room, 23}
[6] \textcolor{OrangeRed}{(User-provided) 3 uses}
\textcolor{RoyalBlue}{(Image caption) Meeting room 03, blinds.}
\end{tabularx}
\end{subtable}
\caption{Examples of collected data in the office inspection tasks showing movement trajectories (gray lines), positions of identified landmarks (blue points), some taken photos (machine-translated), user-provided texts (red texts), and image captions generated from the photos (blue texts). Each trajectory has its own origin $(0, 0)$ and potentially varying true-north; this requires our alignment algorithm for creating a unified semantic landmark map.}
\Description{The figure displays three trajectory plots, each illustrating a path taken through a spatial environment with specific points of interest marked along the way. The first plot shows a trajectory with three labeled points [1], [2], and [3], each connected to images of door signs indicating room numbers: "22," "19," and another room number. The path follows a curved route with these points positioned sequentially along the trajectory. The second plot features three labeled points [1], [2], and [3], linked to images of door signs displaying room numbers: "16," "26," and another room number. The trajectory in this plot is more complex, with multiple curves and intersections. The third plot includes six labeled points [1], [2], [3], [4], [5], and [6], each associated with images of door signs showing room numbers: "16," "26," and others. The path in this plot demonstrates a winding trajectory with several loops and turns.}
\label{fig:office_inputs}
\end{figure}

\begin{figure}[t]
\centering
\begin{subfigure}[b]{0.33\linewidth}
\includegraphics[width=\textwidth]{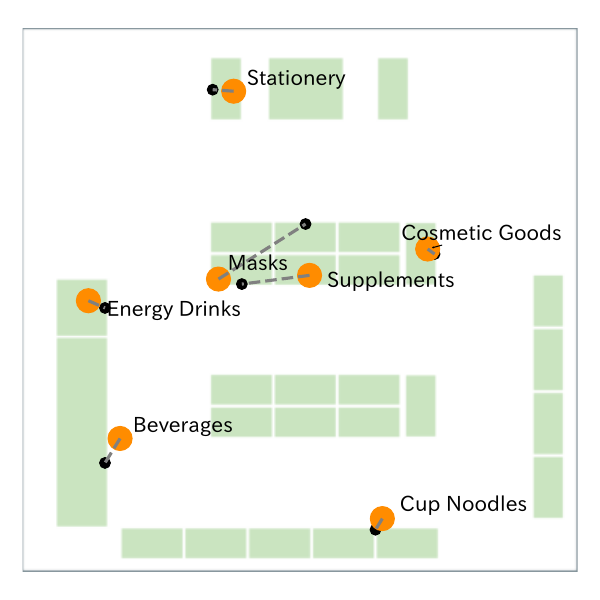}
\caption{\#records = 3, user-provided text}
\end{subfigure}
\begin{subfigure}[b]{0.33\linewidth}
\includegraphics[width=\textwidth]{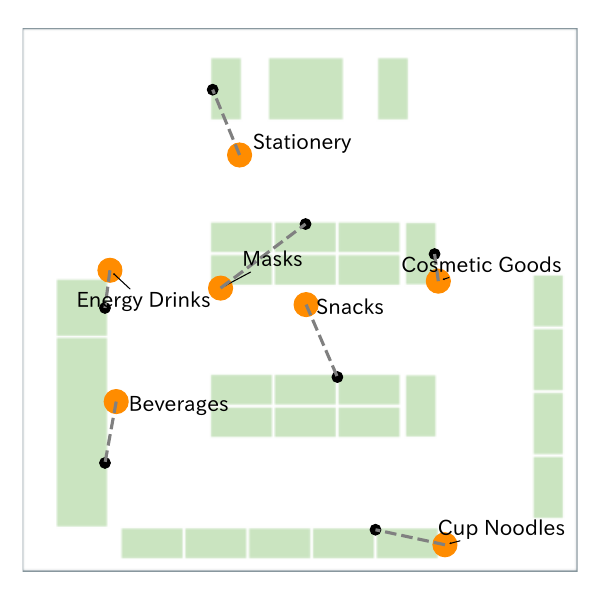}
\caption{\#records = 3, image captioning}
\end{subfigure}
\begin{subfigure}[b]{0.33\linewidth}
\includegraphics[width=\textwidth]{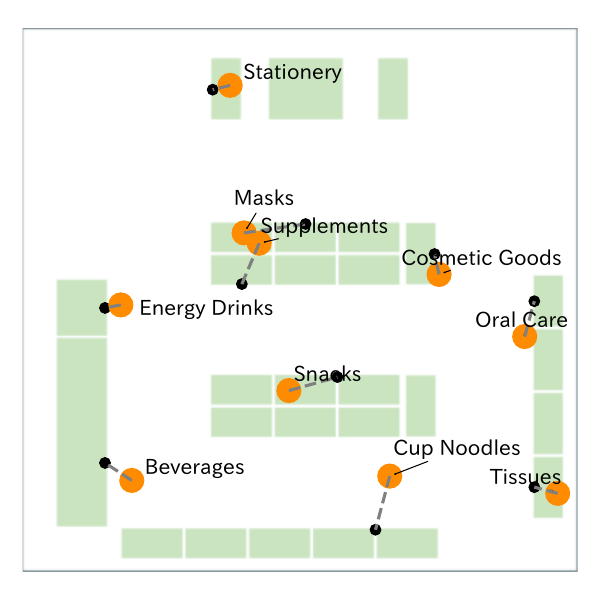}
\caption{\#records=11, user-provided text}
\end{subfigure}
\begin{subtable}[h]{\linewidth}
\centering
\tiny
    \begin{tabularx}{\linewidth}{lcX}
    \\
    \toprule
    Landmark name & Position & Product availability \\
    \midrule
    Beverages & x=1.6m, y=7.3m & Calpis Water 500 (21 bottles) (2025-01-17 11:36:37), Healthy Mineral Barley Tea (Plenty available) (2025-01-17 10:41:49), Super Carbonated Water Lemon (Many) (2025-01-15 12:08:35), Welch's 1 Bunch of Grapes (A lot) (2025-01-17 10:44:25), Cheerio Lemon (Many remaining) (2025-01-15 12:04:49), Ito En Healthy Mineral Barley Tea (Many) (2025-01-15 11:01:17), Barley Tea (Many) (2025-01-15 10:20:53), Iyemon Rich Flavor (5 pieces) (2025-01-15 12:01:54)  \\
    Cosmetic Goods & x=6.3m, y=3.9m & Clip-in Bang Curler (None) (2025-01-15 12:12:12), B Lash W Eyebrow Gel (3 remaining) (2025-01-15 11:04:30), Perfect EX Mascara Pure Brown (1 remaining) (2025-01-15 10:53:53), Magic Concealer Pink Beige (1 remaining) (2025-01-15 10:08:21), W Eyebrow Gel (1 remaining) (2025-01-15 10:14:20), Dry Skin Moisturizing Skin Care Base (3) (2025-01-17 10:44:25), Shonan Beauty Eyelash Serum (2 available) (2025-01-17 11:43:13), Eye Liquid Skin Film (1 remaining) (2025-01-17 10:19:15), Silky Liquid Eyeliner Black (1 remaining) (2025-01-17 10:23:19), Tweezers Scissors (1 remaining) (2025-01-15 10:18:24), Eyelash Serum (2) (2025-01-15 10:20:53)  \\
    Cup Noodles & x=5.6m, y=7.0m & Nissin Cup Noodle Curry (Many) (2025-01-15 12:12:12), Shin Ramyeon (Plenty available) (2025-01-17 10:41:49), Nissin Cup Noodle Mini (Many) (2025-01-15 11:01:17), Maruchan Cup Wonton Soy Sauce (10 remaining) (2025-01-17 10:23:19), Soup Harusame Wonton (5 remaining) (2025-01-15 11:32:02), Cup Noodle Curry (Many remaining) (2025-01-15 10:18:24)  \\
    Energy Drinks & x=1.5m, y=4.1m & Monster Zero Sugar (Many) (2025-01-15 10:08:21), Monster Energy Pipeline Punch (More than 20 available) (2025-01-17 11:43:13), Lipovitan D (3 remaining) (2025-01-15 12:04:49), Monster Green (Many remaining) (2025-01-15 11:28:33), Black Vinegar Energy Drink (8 remaining) (2025-01-17 10:23:19), Red Bull (A lot remaining) (2025-01-15 11:32:02), Chocola BB Royal (4) (2025-01-15 10:20:53), Red Bull (Many remaining) (2025-01-15 12:01:54)  \\
    Masks & x=3.4m, y=3.8m & Alcohol-free Disinfecting Outdoor (2 pieces) (2025-01-17 11:36:37), Mask Code 3D Mask (2) (2025-01-15 12:12:12), Nodo Nuru Wet Mask for Sleep (3 remaining) (2025-01-15 11:04:30), Mask Code 3D Mask Greige (Available) (2025-01-15 10:53:53), Unicharm Ultra Comfortable Mask (4 available) (2025-01-17 10:41:49), Ultra Comfortable Mask Breathable Clear Normal (4) (2025-01-15 12:08:35), Fitti 7 Days Mask EX Normal (3) (2025-01-17 10:44:25), Unicharm Ultra Comfortable Pleated Type 30 sheets (3 available) (2025-01-17 11:43:13), Mask Mix Mask (1 remaining) (2025-01-15 12:04:49), Mask Code 3D Mask Greige (2 remaining) (2025-01-15 11:28:33), Ultra Comfortable SMART COLOR Mask (1 remaining) (2025-01-17 10:19:15), Iris Nano Air Mask (5 remaining) (2025-01-15 11:01:17), Nano Air Mask (5 remaining) (2025-01-15 11:32:02), Ultra Comfortable Smart Color Mask (1 remaining) (2025-01-15 10:11:34), Mask Code 3D Mask (1 remaining) (2025-01-15 10:18:24), Small Face Mask (3) (2025-01-15 10:20:53), Smart Color Mask (1 remaining) (2025-01-15 12:01:54)  \\
    Oral Care & x=7.9m, y=4.9m & Regular Toothbrush 10 pieces (1 remaining) (2025-01-15 10:53:53), OCH Tune Toothbrush (1 remaining) (2025-01-15 10:14:20), Shumitect Periodontal Care (2 remaining) (2025-01-17 10:19:15), Floss Pick (8 remaining) (2025-01-15 10:11:34)  \\
    Snacks & x=4.8m, y=5.3m & Ottotto Thin Salt Flavor (5) (2025-01-17 11:36:37), Bourbon Petit Chocolate Chip (None) (2025-01-15 12:12:12), Chip Star S Thin Salt (6) (2025-01-15 12:08:35), Sapporo Potato Veggie Bits (About 10 available) (2025-01-17 11:43:13), Chip Star Thin Salt (6 remaining) (2025-01-15 11:28:33), Snap Pea Snack (Many remaining) (2025-01-15 10:18:24)  \\
    Stationery & x=3.6m, y=0.7m & Tama Electronics CL Cable 2m (2 remaining) (2025-01-15 11:04:30), Tombow Eraser Mono (3) (2025-01-17 10:44:25), Lightning Cable 50 cm (1 remaining) (2025-01-15 12:04:49), Battery-powered Charger (1 remaining) (2025-01-17 10:23:19)  \\
    Supplements & x=3.2m, y=3.3m & DHC Blueberry 20 days 40 capsules (1 remaining) (2025-01-15 11:04:30), Alinamin V (2) (2025-01-15 12:08:35), Turmeric Power Turmeric Extract (3) (2025-01-15 10:14:20), Turmeric Power (3 remaining) (2025-01-17 10:19:15), DHC Multivitamin (1 remaining) (2025-01-15 11:32:02), Gummy Supplement (2 remaining) (2025-01-15 10:11:34), Ulco Supplement (2 remaining) (2025-01-15 10:20:53)  \\
    Tissues & x=8.2m, y=6.8m & Elleair Pocket Tissue (1) (2025-01-15 12:12:12), Elleair + Water Pocket 14W4P (2 remaining) (2025-01-15 11:04:30), Flushable Pocket Tissue (1 remaining) (2025-01-15 10:08:21), Silver Ion Wet Wipes Tissue (6) (2025-01-15 10:14:20), Hana Celeb Pocket Tissue (1 remaining) (2025-01-15 11:28:33), Hana Celeb Pocket Tissue (1 remaining) (2025-01-15 11:01:17) \\
    \bottomrule
    \end{tabularx}
    \caption{Product availability (\# records = 20)}
\end{subtable}
\caption{Product availability maps generated from restocking checklist tasks. (a-c) Identified landmarks are annotated with orange circles, while their corresponding ground-truth positions are marked with small black circles. Green rectangles indicate the rooms or shop shelves in the environments. (d) Product availability aggregated with shelf locations.}
\Description{There are two rows of floor plan visualizations. The first row consists of two visualizations; the underlying floor plans are the same and it is for the office environment used in the study. The annotations depict the difference in positional accuracy of the localized landmarks depending on the number of trajectories provided by the users. The second row has three visualizations. For all three, the underlying floor plan is for the indoor environment used for the retail store scenario. Like the visualization above, the annotations show the difference in positional errors.}
\label{fig:qualitative_store}
\vspace{.5em}
\end{figure}

\begin{figure}[t]
\centering
\begin{subfigure}[b]{0.49\linewidth}
\includegraphics[width=\textwidth]{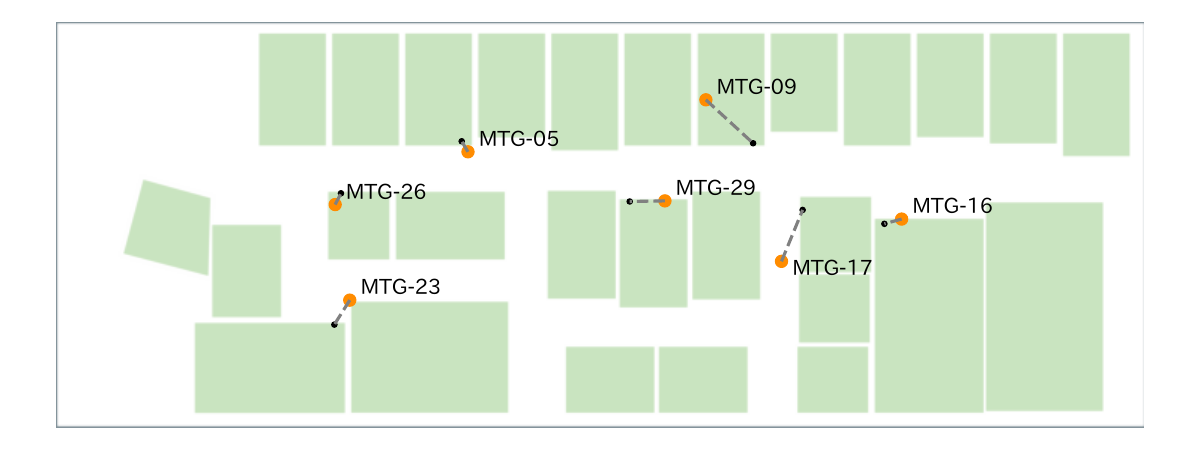}
\caption{\#records = 3, user-provided text}
\end{subfigure}
\begin{subfigure}[b]{0.49\linewidth}
\includegraphics[width=\textwidth]{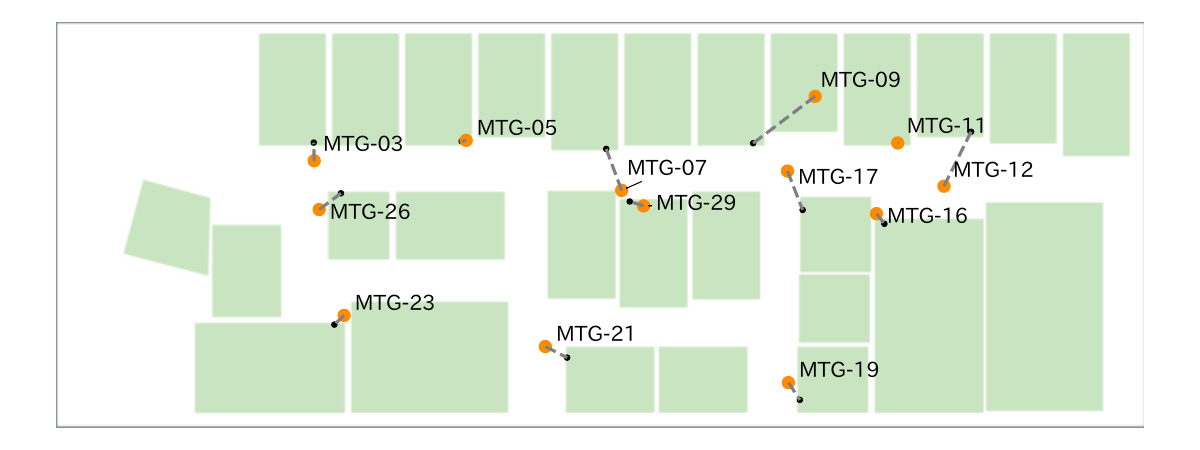}
\caption{\#records = 11, user-provided text}
\end{subfigure}
\begin{subtable}[h]{0.99\linewidth}
\centering
\tiny
    \begin{tabularx}{\linewidth}{lcX}
    \\
    \toprule
    Landmark name & Position & Usage history \\
    \midrule
    MTG-03 & x=13.4m, y=6.4m & occupied (2025-01-08 10:36:47), occupied (2025-01-08 10:42:24), available (2025-01-09 11:16:16), occupied (2025-01-09 13:01:45), occupied (2025-01-09 10:36:28), in use (2025-01-09 14:05:46)  \\
    MTG-05 & x=22.6m, y=6.2m & occupied (2025-01-08 10:45:13), available (2025-01-09 11:16:16), available (2025-01-09 13:01:45), available (2025-01-09 11:33:20), available (2025-01-09 11:09:34), available (2025-01-09 11:12:43), available (2025-01-09 11:06:31), vacant room (2025-01-09 10:45:36)  \\
    MTG-07 & x=30.2m, y=6.1m & occupied (2025-01-08 10:36:47), occupied (2025-01-08 10:42:24), occupied (2025-01-09 11:29:31), occupied (2025-01-20 10:50:32), occupied (2025-01-09 11:09:34), occupied (2025-01-09 11:06:31), available (2025-01-09 12:47:18)  \\
    MTG-09 & x=37.0m, y=7.9m & available (2025-01-08 10:45:13), available (2025-01-08 10:36:47), occupied (2025-01-09 11:29:31), occupied (2025-01-09 11:16:16), available (2025-01-09 12:47:18), available (2025-01-09 12:42:54), vacant room (2025-01-09 12:59:07)  \\
    MTG-11 & x=44.7m, y=6.4m & occupied (2025-01-08 10:36:47), occupied (2025-01-08 10:42:24), vacant (2025-01-09 11:16:16), available (2025-01-09 11:33:20), occupied (2025-01-08 10:33:56), available (2025-01-09 11:09:34), vacant room (2025-01-09 14:05:46), available (2025-01-09 11:06:31), available (2025-01-09 10:45:36), vacant room (2025-01-09 12:59:07)  \\
    MTG-12 & x=48.9m, y=7.2m & occupied (2025-01-09 11:02:11), occupied (2025-01-20 10:50:32), occupied (2025-01-09 11:33:20), occupied (2025-01-09 11:09:34), occupied (2025-01-09 12:47:18), occupied (2025-01-09 12:42:54)  \\
    MTG-16 & x=46.6m, y=10.6m & occupied (2025-01-08 10:45:13), available (2025-01-09 10:40:44), available (2025-01-09 10:36:28), available (2025-01-20 10:50:32), available (2025-01-09 11:33:20), occupied (2025-01-09 12:47:18), available (2025-01-09 12:42:54)  \\
    MTG-17 & x=40.8m, y=12.0m & available (2025-01-09 10:40:44), occupied (2025-01-09 11:02:11), available (2025-01-20 10:50:32), occupied (2025-01-09 11:09:34), occupied (2025-01-09 11:12:43), available (2025-01-09 12:42:54), vacant room (2025-01-09 12:59:07)  \\
    MTG-19 & x=40.5m, y=17.6m & available (2025-01-09 11:29:31), available (2025-01-09 11:16:16), available (2025-01-09 11:33:20), vacant room (2025-01-09 14:05:46)  \\
    MTG-21 & x=26.5m, y=19.6m & available (2025-01-09 13:01:45), occupied (2025-01-09 10:36:28), occupied (2025-01-09 11:12:43), in use (2025-01-09 10:45:36)  \\
    MTG-22 & x=26.5m, y=15.5m & available (2025-01-09 11:02:11), available (2025-01-09 11:06:31), available (2025-01-09 12:47:18), vacant room (2025-01-09 10:45:36)  \\
    MTG-23 & x=14.7m, y=15.4m & occupied (2025-01-08 10:45:13), available (2025-01-09 10:40:44), occupied (2025-01-08 10:42:24), vacant (2025-01-09 11:16:16), available (2025-01-09 10:36:28), available (2025-01-09 11:12:43), vacant room (2025-01-09 14:05:46)  \\
    MTG-26 & x=14.8m, y=10.9m & occupied (2025-01-08 10:45:13), occupied (2025-01-09 10:40:44), occupied (2025-01-08 10:42:24), occupied (2025-01-09 13:01:45), occupied (2025-01-09 10:36:28), occupied (2025-01-08 10:33:56), occupied (2025-01-09 11:12:43), in meeting (2025-01-09 10:45:36)  \\
    MTG-29 & x=33.3m, y=8.5m & occupied (2025-01-08 10:45:13), occupied (2025-01-09 10:40:44), occupied (2025-01-20 10:50:32), occupied (2025-01-08 10:33:56), available (2025-01-09 12:47:18), in meeting (2025-01-09 10:45:36)  \\
    \bottomrule
    \end{tabularx}
    \caption{Room usage map (\# records = 20)}
\end{subtable}
\caption{Room usage maps generated from office inspection tasks. (a-c) Identified landmarks are annotated with orange circles, while their corresponding ground-truth positions are marked with small black circles. Green rectangles indicate the rooms or shop shelves in the environments. (d) Room usage history aggregated with room locations.}
\Description{There are two rows of floor plan visualizations. The first row consists of two visualizations; the underlying floor plans are the same and it is for the office environment used in the study. The annotations depict the difference in positional accuracy of the localized landmarks depending on the number of trajectories provided by the users. The second row has three visualizations. For all three, the underlying floor plan is for the indoor environment used for the retail store scenario. Like the visualization above, the annotations show the difference in positional errors.}
\label{fig:qualitative_office}
\end{figure}

\paragraph{Data and Map Examples}
Fig.~\ref{fig:store_inputs} and Fig.~\ref{fig:office_inputs} showcase selected examples of collected data. For the restocking checklist task in Fig.~\ref{fig:store_inputs}, both participants' texts and image captions provided information on various products available on shelves, \eg, \textit{``My lash mascara has 3 left,'' ``Mitsubishi Jet Stream 05 red,''}\footnote{Participants' descriptions are all machine-translated into English from their native languages and may not convey the original nuance.} although image captions were sometimes less accurate due to misleading package appearances. For the office inspection tasks in Fig.~\ref{fig:office_inputs}, participants described room availability in various ways, \eg, \textit{``Room 22 available,'' ``No utilization of 19.''} Image captions generated from photos successfully contained room numbers displayed on room doors (\eg, \textit{``Meeting Room 22, Door''}). Fig.~\ref{fig:qualitative_store} visualizes examples of generated semantic landmark maps for restocking checklist tasks. Our system constructed product availability maps that can inform users about the availability of various product items for each shelf with last confirmation times. Fig.~\ref{fig:qualitative_office} shows examples of landmark maps for office inspection tasks, which describe the locations of rooms with their recent usage history.

\begin{figure}[t]
\centering
\begin{subfigure}[b]{0.49\linewidth}
\includegraphics[width=\textwidth]{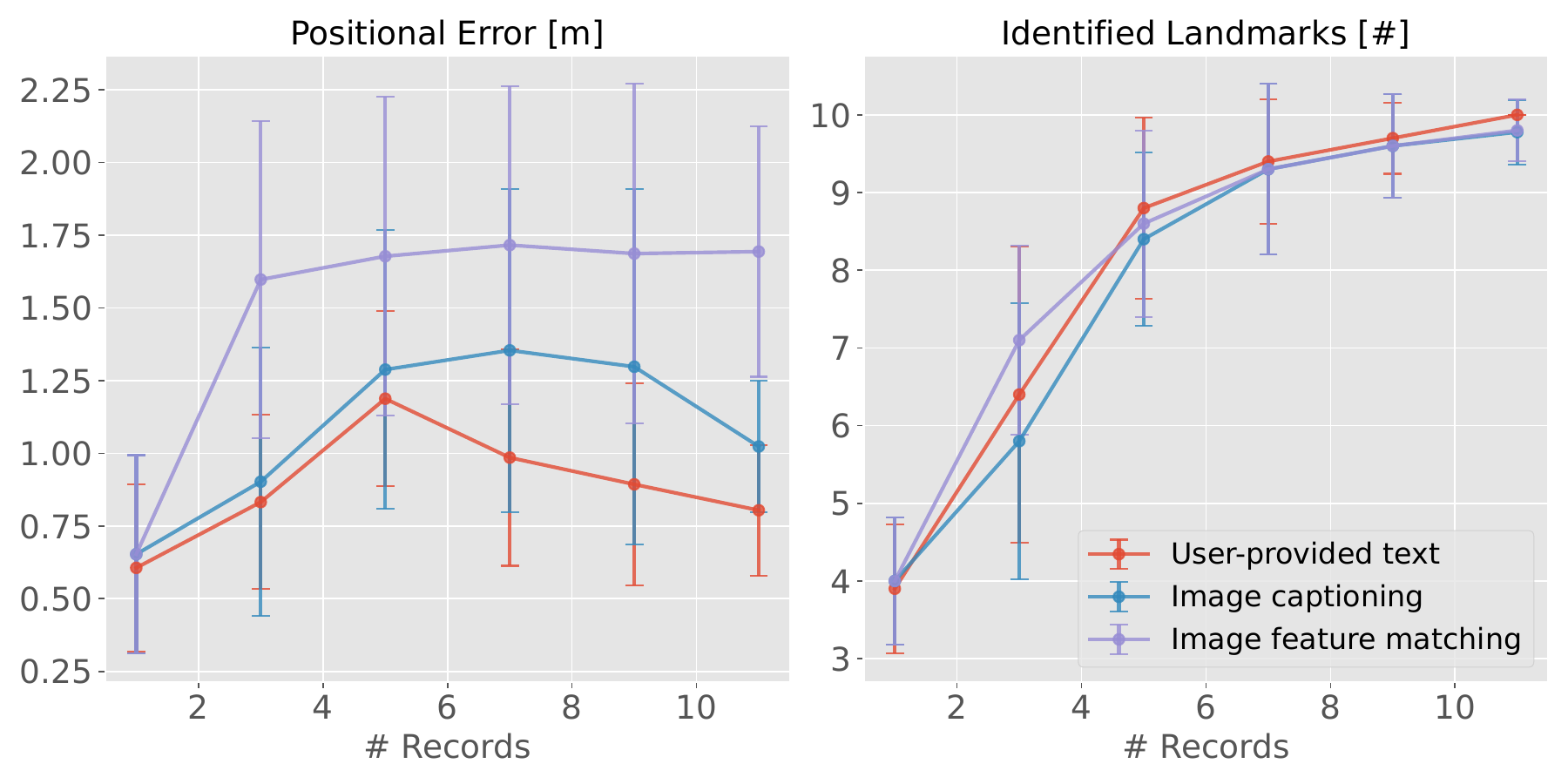}
\caption{Restocking checklist tasks}
\end{subfigure}
\begin{subfigure}[b]{0.49\linewidth}
\includegraphics[width=\textwidth]{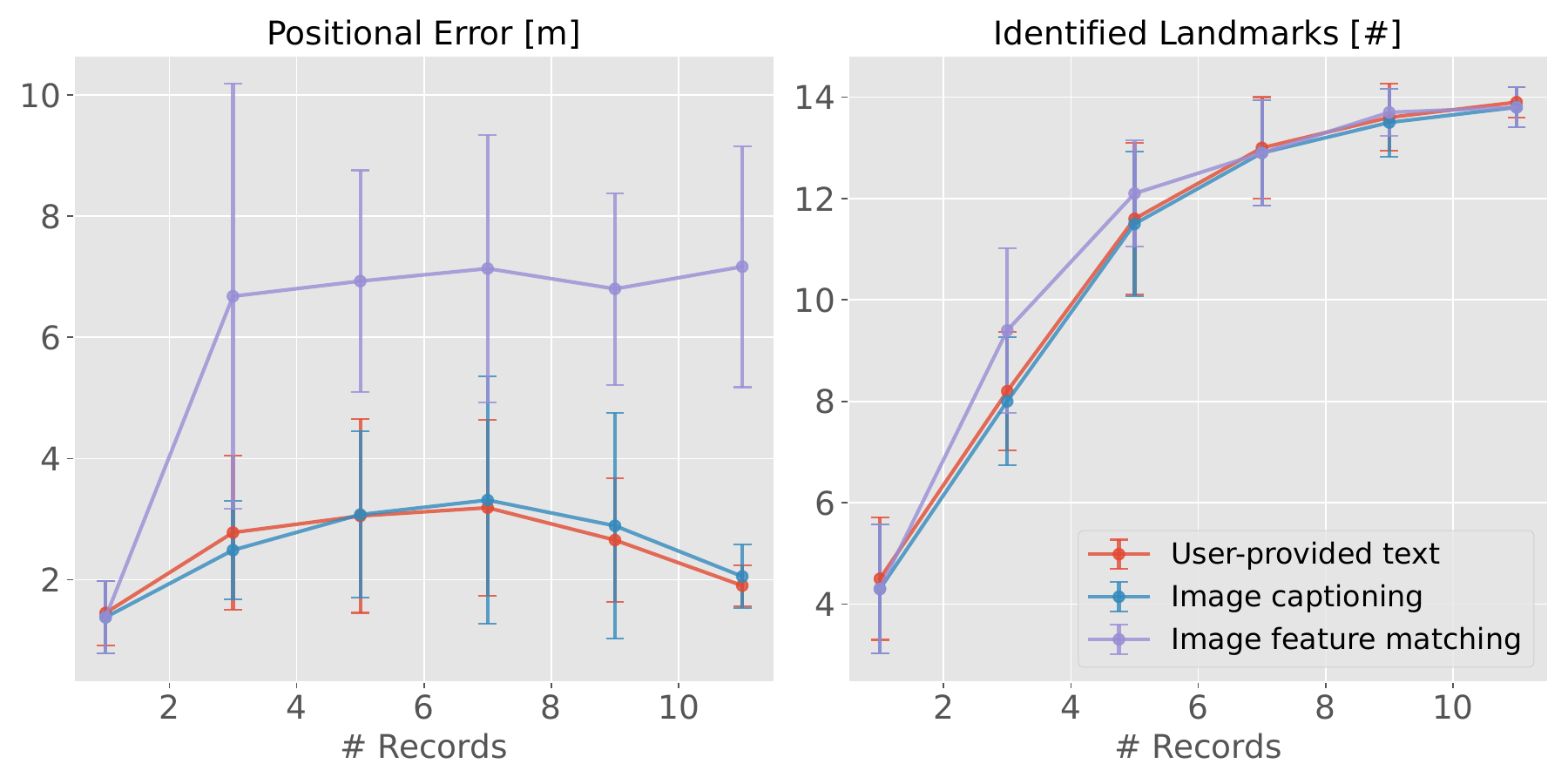}
\caption{Office inspection tasks}
\end{subfigure}
\caption{Quantitative evaluation. We evaluate the positional error (left) between identified and ground-truth landmarks, as well as the number of identified landmarks (right) while changing the number of records to aggregate.}
\Description{There are four line charts placed in one row. The first two charts represent the evolution of positional error (in meters) and the number of identified landmarks for the office inspection scenario. Each chart has three lines, corresponding to the three algorithmic conditions: ``user-provided text'', ``image captioning'', and ``image feature matching''. The two remaining charts are similar to the first pair, but is for the restocking checklist scenario.}
\label{fig:plot}
\end{figure}

\paragraph{Positional Error}
In Fig.~\ref{fig:plot}, we found a unique tendency in positional errors that increased first and then decreased with an increasing number of records. This can be intuitively interpreted by the possible number of records used to derive the position for each identified landmark. If the number of records is relatively limited for a large number of identified landmarks, the performance can suffer more from positioning errors in each record. Compared to user-provided texts, image captioning worked comparably well for the office inspection tasks but did not perform as well for the restocking checklist tasks. In the office space, each room door displays its room number, which can be easily recognized by the LLM. In contrast, it was not always easy to perfectly recognize product items from photos in the retail store, leading to landmark misidentification and mapping inaccuracy. Note also that image captioning incurs extra communication and API costs, as well as potential privacy concerns if the system is running during busy business hours, compared to the proposed method. As shown in Fig.~\ref{fig:store_inputs} and \ref{fig:office_inputs}, the visual appearances of photos are diverse in the retail store and significantly similar in the office, which both made it difficult for the image-feature matching to stably compute the relatedness between landmarks.

\paragraph{Landmark Coverage}
As also shown in Fig.~\ref{fig:plot}, the coverage (\ie, number of identified landmarks) monotonically increases with the number of records, arriving at nearly 100\% identification when data from 11 users are aggregated. Note that the image-feature matching baseline tends to identify more landmarks in office inspection tasks because image features are similar between landmarks (\ie, room doors), which nonetheless resulted in incorrect aggregation as shown by its high positional error.

\begin{figure}[t]
\centering
\begin{subfigure}[b]{0.49\linewidth}
\includegraphics[width=\textwidth]{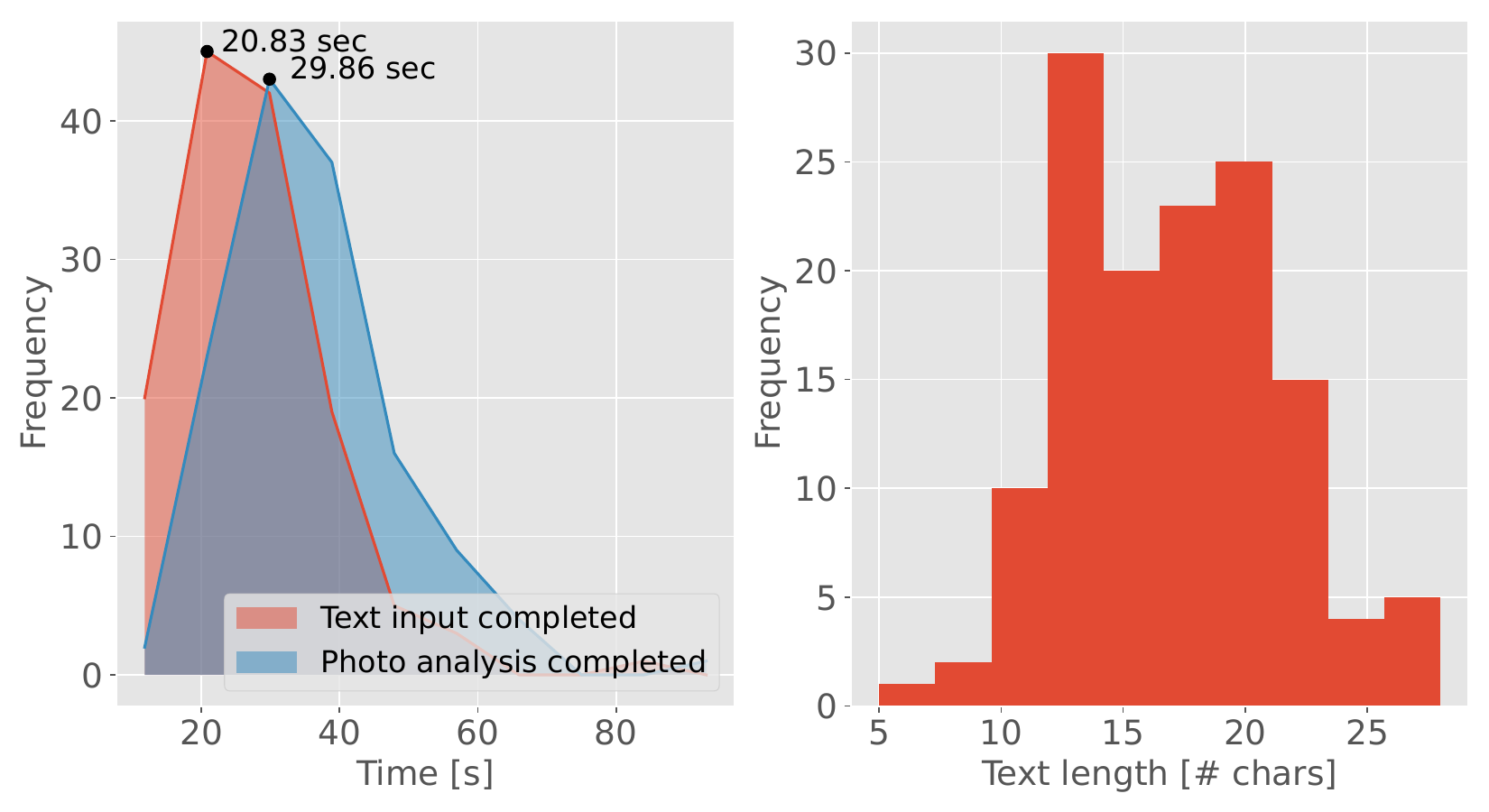}
\caption{Restocking checklist tasks}
\end{subfigure}
\begin{subfigure}[b]{0.49\linewidth}
\includegraphics[width=\textwidth]{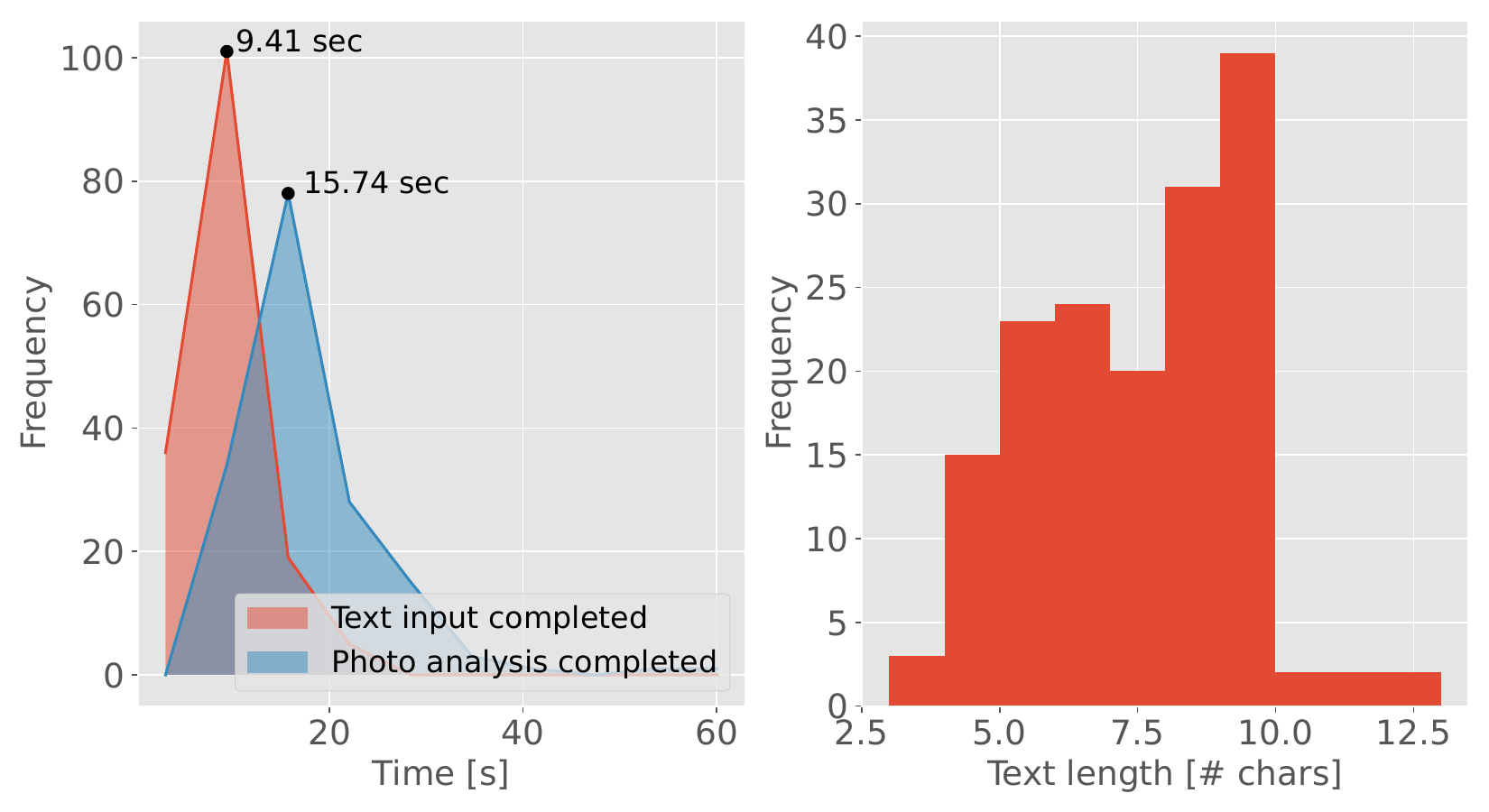}
\caption{Office inspection tasks}
\end{subfigure}
\caption{Time Analysis. We visualized the histograms for times to complete text entries (left: orange) and those to complete photo taking and analysis (left: blue) after arriving at the landmark. Moreover, we obtained another set of histograms of text lengths (number of characters of original text inputs) for respective tasks (right).}
\Description{The figure presents two sets of charts labeled (a) Restocking checklist tasks and (b) Office inspection tasks, each consisting of two charts. For the restocking checklist tasks, the first chart is a frequency distribution of task completion times, with red and blue shaded areas representing the completion of text input and photo analysis, respectively, marked by specific time points (20.83 sec for text input and 29.86 sec for photo analysis). The second chart shows a histogram of text lengths in characters, ranging from about 5 to 30 characters. Similarly, for the office inspection tasks, the first chart displays a frequency distribution of completion times, with text input completed at 9.41 seconds and photo analysis at 15.74 seconds. The second chart is a histogram of text lengths, with a range from approximately 2.5 to 12.5 characters. Both sets of charts visually compare the timing and text length characteristics of the respective tasks.}
\label{fig:time_analysis}
\end{figure}

\begin{figure}[t]
\centering
\includegraphics[width=\textwidth]{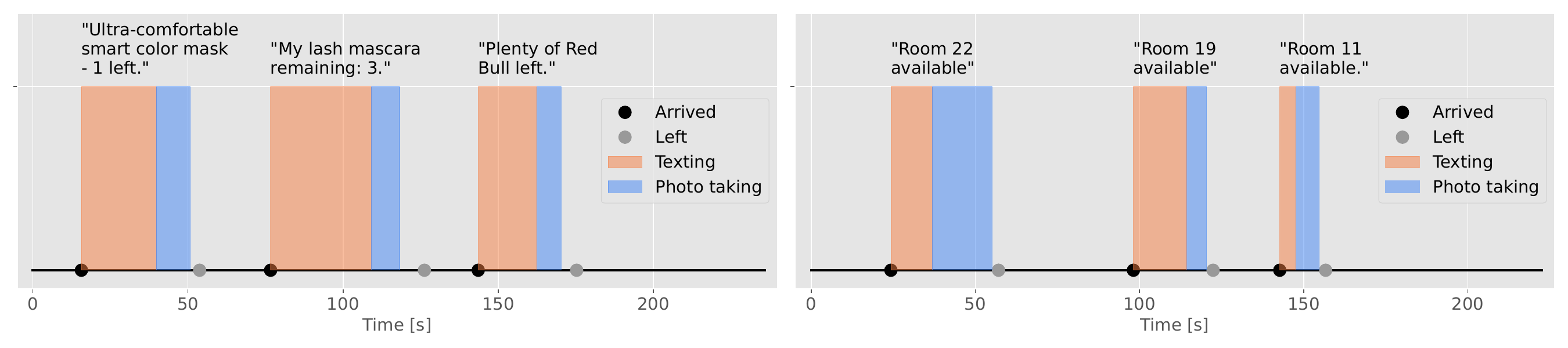}
\caption{Examples of task progress timelines. The moments of arriving at and leaving from landmarks are marked with black and gray circles, respectively. The time intervals where text inputs and photo taking/analysis are happening are described by orange and blue rectangles, respectively.}
\Description{The figure consists of two timeline charts placed in one row. Each chart represents a sequence of events over time, with specific activities marked along the timeline. The first chart includes events related to product availability notifications, such as ``Ultra-comfortable smart color mask - 1 left,'' ``My lash mascara remaining: 3,'' and ``Plenty of Red Bull left.'' The second chart focuses on room availability notifications, with events like ``Room 22 available,'' ``Room 19 available,'' and ``Room 11 available.'' In both charts, events are marked with black dots for arrival and gray dots for departure. Additionally, activities are color-coded: orange for texting and blue for photo-taking.}
\label{fig:time_example}
\vspace{1em}
\end{figure}

\paragraph{User Interaction and Time Analysis}
Overall, participants were able to fully understand how to use our data-collection application with approximately three minutes of explanation and could input text without confusion during the study. Because all participants were familiar with the environment, there was less confusion about where to go next while on the move. However, the distance from which they entered text relative to landmarks (\eg, room doors and product shelves) varied among participants, and some participants began inputting text as soon as landmarks became visible. Fig.~\ref{fig:time_analysis} shows statistics on how participants interacted with our interface, while Fig.~\ref{fig:time_example} elaborates on several concrete examples. Text inputs required 20.83 seconds on average for restocking checklist tasks and 9.41 seconds for office inspection tasks. This difference came from text length, where the former was twice as long as the latter. Additional time overhead for taking and processing photos was 6-9 seconds on average, which was more significant in office inspection tasks with shorter text inputs as also shown in Fig.~\ref{fig:time_example}. Many participants conducted the tasks while paying attention to their surroundings so as not to disturb other people in the environments. In particular, we observed that participants sometimes had to pause taking photos to wait for people crossing in front of them. Despite no specific instructions, all participants tended to stay at the landmark until they were sure the photos had been processed correctly by the LLM and captions had been provided. Some were simply interested in LLMs behaving intelligently, while some others were trying to make sure they produced accurate outputs while imagining the purpose of the task.

\subsection{Discussion}

Overall, our results confirmed that users can provide free text inputs with sufficiently detailed information for our LLM module to identify landmark names during their tasks. As shown in Fig.~\ref{fig:time_analysis}, by extracting landmark-related features from user-provided texts during primary tasks such as creating restocking checklists or office room inspection, the overall time spent at each landmark can be reduced by up to 60\% compared to when additionally asking users to take photos and confirm captioning results solely for map creation. This demonstrates the benefit of creating a map as a by-product, particularly for retailers who seek operational efficiency in the face of personnel shortages and limited business hours as confirmed in Sec.~\ref{subsec:needs}; tasks like shelf checking and restocking constitute the vast majority of non-cashier work time (up to 90\% reported by one interviewee), making this an ideal workflow to leverage for data collection without adding significant overhead.

We note, however, that in the retail checklist task, some product classifications challenged the LLM. For example, distinguishing between similar categories was difficult; vitamin B drinks marketed for fatigue recovery (\eg, \textit{`Alinamin V'}) were sometimes categorized as `Supplements' and other times as `Energy Drinks' depending on the specific user input, highlighting the challenge of mapping free-form text to fixed categories. We believe we could mitigate the name misidentification problem in a practical setting. For example, retailers could provide their existing product catalogs with precise categorizations to the LLM, although this would increase API costs. Interestingly, even with the presence of name misidentification, we did not see devastating consequences for the accuracy of locating landmarks (\ie, shelves) when there were enough records. This demonstrates the robustness of our optimization-based method for identifying landmark locations. Our system provided accurate landmark maps with an average positional error of 0.8 m for 10 out of 10 landmarks in a 70 m$^2$ retail store and 1.9 m for an average of 13.9 out of 14 landmarks in a 300 m$^2$ office by aggregating 11 records. This was achieved despite each record containing only 3-6 landmarks, with some mistakes in landmark identification and partially inaccurate positioning due to IMU sensor noise. Positional errors further improved to 0.63 m for 20 records and 0.56 m for 30 records in the retail store, and 1.4 m for 20 records and 1.2 m for 30 records in the office. The acceptable level of positional accuracy would vary by application, but for use cases such as indoor landmark-based navigation, we believe these positional errors should be acceptable. For applications that demand high positional precision, we recommend collecting as many records as possible, particularly when the potential number of identifiable landmarks is large. Our simulation experiment in the following section investigates the effects of the number of records and the number of landmarks for each record in more detail.

Beyond mapping the physical locations of shelves and rooms, our method collected dynamic semantic information that updates over time. As presented in Fig.~\ref{fig:store_inputs} and Fig.~\ref{fig:office_inputs}, the user-provided data in the current user study included detailed product inventory levels and real-time room occupancy status, which were aggregated to generate the product availability maps in Fig.~\ref{fig:qualitative_store} and room usage maps in Fig.~\ref{fig:qualitative_office}, respectively. Due to the flexible nature of the input format (\ie, free text), our approach is very versatile in collecting different types of dynamic information in conjunction with their locations. For example, it is straightforward to extend our approach to accept tabular databases already used in regular tasks. Though we did not formally evaluate the difference in quality between human-generated data and LLM-generated data, anecdotally, human-provided data contained richer semantic information beyond what was used for mapping. For example, it is difficult even for present state-of-the-art LLMs to precisely count the number of remaining items on a shelf solely from a single photo in which items are occluded by each other.

\looseness=-1
The collected data can enable the development of location-based services for employees or customers. As a proof-of-concept, we implemented an in-store/office chatbot as shown in Tab.~\ref{tab:chat_bot}. Here, the generated maps capture physical locations and dynamic product availability or room usage history with timestamps. These data can serve as a knowledge base for LLMs to assist users in locating desired items or rooms. Future work plans to investigate how chatbots enabled by landmark maps could improve user experience and optimize business operations.

\begin{table}[t]
    \centering
    \caption{In-store/office chatbot. We ask an LLM (GPT-4o) the questions in the first column, along with semantic information (product availability in Fig.~\ref{fig:qualitative_store} or room usage history in Fig.~\ref{fig:qualitative_office}) to get answers in the second column.}
     \scriptsize
    \begin{tabularx}{\linewidth}{lX}
    \toprule
         Questions & Chatbot's answers \\
         \midrule
        Are there any items that need to be restocked? & 
        Here is a list of those items along with their positions:

**Cosmetic Goods:**
1. Clip-in Bang Curler - None available (Position: x=6.3m, y=3.9m)
2. Perfect Ex Mascara Pure Brown - 1 remaining (Position: x=6.3m, y=3.9m)
3. Magic Concealer Pink Beige - 1 remaining (Position: x=6.3m, y=3.9m)
4. W Eyebrow Gel - 1 remaining (Position: x=6.3m, y=3.9m)
5. Eye Liquid Skin Film - 1 remaining (Position: x=6.3m, y=3.9m)
6. Silky Liquid Eyeliner Black - 1 remaining (Position: x=6.3m, y=3.9m)
7. Tweezers Scissors - 1 remaining (Position: x=6.3m, y=3.9m)

**Energy Drinks:**
1. Lipovitan D - 3 remaining (Position: x=1.5m, y=4.1m)
2. Chocola BB Royal - 4 pieces (Position: x=1.5m, y=4.1m)

**Masks:**
1. Mask Mix Mask - 1 remaining (Position: x=3.4m, y=3.8m)
2. Ultra Comfortable Smart Color Mask - 1 remaining (Position: x=3.4m, y=3.8m)

**Oral Care:**
1. Regular Toothbrush 10 - 1 remaining (Position: x=7.9m, y=4.9m)
2. OCH Tune Toothbrush - 1 remaining (Position: x=7.9m, y=4.9m)

**Snacks:**
1. Bourbon Puchi Chocolate Chips - None available (Position: x=4.8m, y=5.3m)

**Stationery:**
1. Lightning Cable 50 cm - 1 remaining (Position: x=3.6m, y=0.7m)
2. Battery-Powered Charger - 1 remaining (Position: x=3.6m, y=0.7m)

**Supplements:**
1. DHC Blueberry 20 Days 40 Tablets - 1 remaining (Position: x=3.2m, y=3.3m)
2. DHC Multivitamin - 1 remaining (Position: x=3.2m, y=3.3m)

These items may require immediate attention for restocking.\\
I'm around the stationery aisle. How can I get something to eat? & You can head to the nearby Cup Noodles aisle, which is located at coordinates (x=5.6m, y=7.0m). Here are some options you can find there:

1. Nissin Cup Noodle Curry - Many available
2. Shin Ramyeon - Plenty available
3. Nissin Cup Noodle Mini - Many available
4. Maruchan Cup Wonton Soy Sauce - 10 remaining
5. Soup Harusame Wonton - 5 remaining
After you finish with the Cup Noodles, you might also want to check out the Snacks aisle at (x=4.8m, y=5.3m), where you can find items like:

1. Ottotto Light Salt Flavor - 5 pieces available
2. Chip Star S Light Salt - 6 available
3. Sapporo Potato Crushed Vegetable - About 10 available
4. Snow Pea Snack - Many remaining

Just follow the aisle directions to reach these sections, and you'll be able to grab something to eat quickly!\\
\midrule
I'm around MTG-05. Is there any other room likely to be available? & 
As you're near MTG-05, there are a few rooms nearby that are likely available:

1. **MTG-03** (Position: x=14.3m, y=5.8m) - The room has had a history of being available recently, notably on January 20th.
2. **MTG-11** (Position: x=45.4m, y=7.1m) - This room has several available time slots recorded, including January 9th.
3. **MTG-09** (Position: x=37.2m, y=7.9m) - This room also has available time slots in its history.
4. **MTG-16** (Position: x=45.9m, y=9.5m) - There are multiple instances of availability recently, which may indicate it's a good option.
\\
Which room is used most frequently? &
Based on the occupancy history in the database, the room used most frequently is **MTG-26**, located at (x=14.7m, y=10.6m). It has multiple recorded occupations, including:

- **Occupied** (2025-01-08 01:45:13)
- **Occupied** (2025-01-09 01:40:44)
- **Occupied** (2025-01-08 01:42:24)
- **Occupied** (2025-01-09 04:01:45)
- **Occupied** (2025-01-09 01:36:28)
- **Occupied** (2025-01-08 01:33:56)
- **Occupied** (2025-01-09 02:12:43)

It has a total of **7 occupied instances**, indicating it is the most frequently used room.
\\
         \bottomrule
    \end{tabularx}
    \label{tab:chat_bot}
\end{table}

Finally, while our user study was conducted in actual retail/office environments, there are several factors worth discussing for future work:
\begin{itemize}
    \item \textbf{Duplicated Landmarks:} In some cases, multiple shelves within a store occasionally contain identical items or item categories. For example, besides the designated snack section, the area near the cashier often holds supplementary snack displays. Similar issues may arise in office environments, where multiple places may have the same name without unique IDs (\eg, elevators, gates, or toilets). This `duplicated landmark' problem can lead to users associating the same landmark name with distinct physical locations, potentially confusing the optimization algorithm. We systematically investigate this issue through simulation experiments in the following section.
    \item \textbf{Dynamic Layout Changes:} Though less frequent in office environments, retail stores regularly change their product layouts, such as introducing and removing seasonal item displays. This can lead to the appearance, disappearance, or relocation of landmarks. Although our mapping algorithm does not currently have a mechanism to detect these changes, users can mitigate this issue by refreshing data collection and maps periodically according to their sales plans. The dynamic nature of retail layouts, confirmed in interviews with descriptions of seasonal changes and new product introductions managed at the store level, reinforces the need for the system's ability to handle evolving environments, potentially through periodic data refreshes as suggested by current practices.
    \item \textcolor{black}{\textbf{Potential Extensions:} Integrating the proposed system with other sensor data or algorithms enables interesting extensions. For example, using barometer measurements available in modern smartphones can help users map multi-floor environments. Projecting landmark maps onto existing floor plans would also be useful for navigation applications. Additionally, the technique of analyzing texts by LLMs to serve as anchors for alignment may further enhance other mapping solutions, such as collaborative SLAM~\cite{lajoie2022towards}.} 
\end{itemize}

\section{SIMULATION EXPERIMENT}
\label{sec:simulation}
To comprehensively investigate the performance of our algorithm under various conditions beyond the setup in the user study, we conducted a series of simulation experiments.

\subsection{Simulation Setup}
We developed a simple simulation environment consisting of 30 landmarks at random positions, closely matched to the number of shelves in the retail store used in our user study. We assigned a unique ID (\eg, ``ID-00'', ``ID-01'') to $N'=\lfloor (1-p)\times N\rfloor$ landmarks ($0\leq p \leq 1$), while the remaining $30 - N'$ landmarks were assigned IDs randomly chosen from the set of $N'$ unique IDs already assigned. This created a situation where multiple shelves contained duplicated categories of items. Note that when running our system, we regarded these IDs as identified landmark names, removing the effect of inferences by LLMs.

User data were synthesized by picking a random subset of $K$ landmarks, where $K$ was sampled based on one of the following conditions: 
\begin{itemize}
    \item \textbf{Few}: All data visits $3\sim 6$ landmarks 
    \item \textbf{Many}: All data contains $12 \sim 15$ landmarks
    \item \textbf{Mixed}: 10\% of data has $12 \sim 15$ landmarks, while the remaining data visits only $3\sim 6$ landmarks
\end{itemize}
We then injected zero-mean Gaussian noise, sampled independently for x and y coordinates with standard deviation $\sigma$ (where $\sigma\sim0.5$ based on the user study data), into the landmark positions. Subsequently, we randomly rotated and translated the entire set of positions for each simulated user record so that the first visited landmark was located at the origin ($[0, 0]$). In this way, we simulated users' positional data with unknown initial yaw, random starting positions, and variable positions when looking at each landmark.

With this simulation, we investigated the effect of: 1) sampling conditions for $K$, 2) noise scale $\sigma$, and 3) landmark duplication rate $p$. Each simulation configuration was evaluated five times with different random seeds (\ie, different landmark positions, different landmark assignments for each record) to calculate average scores.

\begin{figure}[t]
\centering
\begin{subfigure}[b]{0.49\linewidth}
\includegraphics[width=\textwidth]{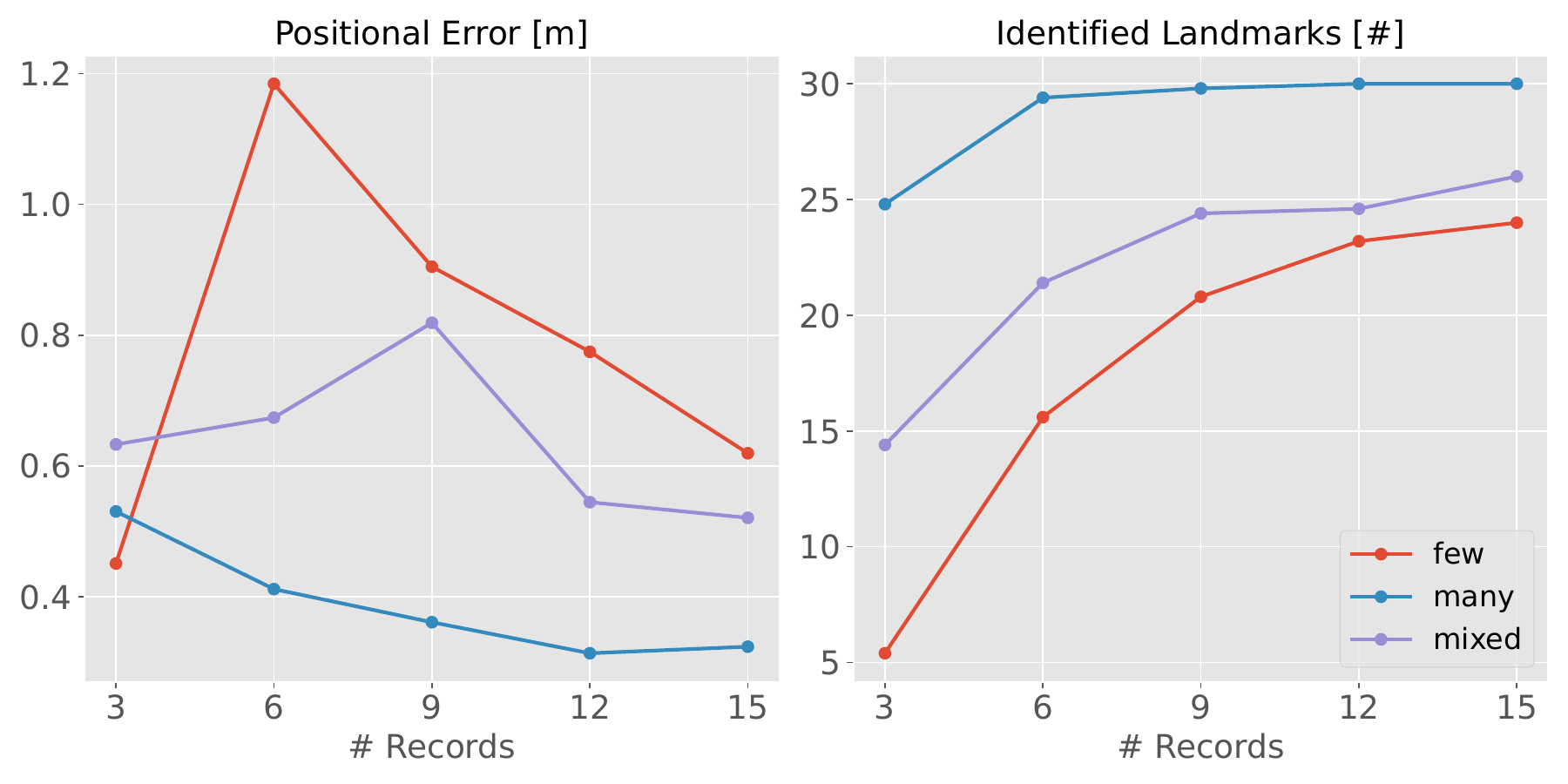}
\caption{Landmark Sampling (few, many, mixed)}
\end{subfigure}
\begin{subfigure}[b]{0.24\linewidth}
\includegraphics[width=\textwidth]{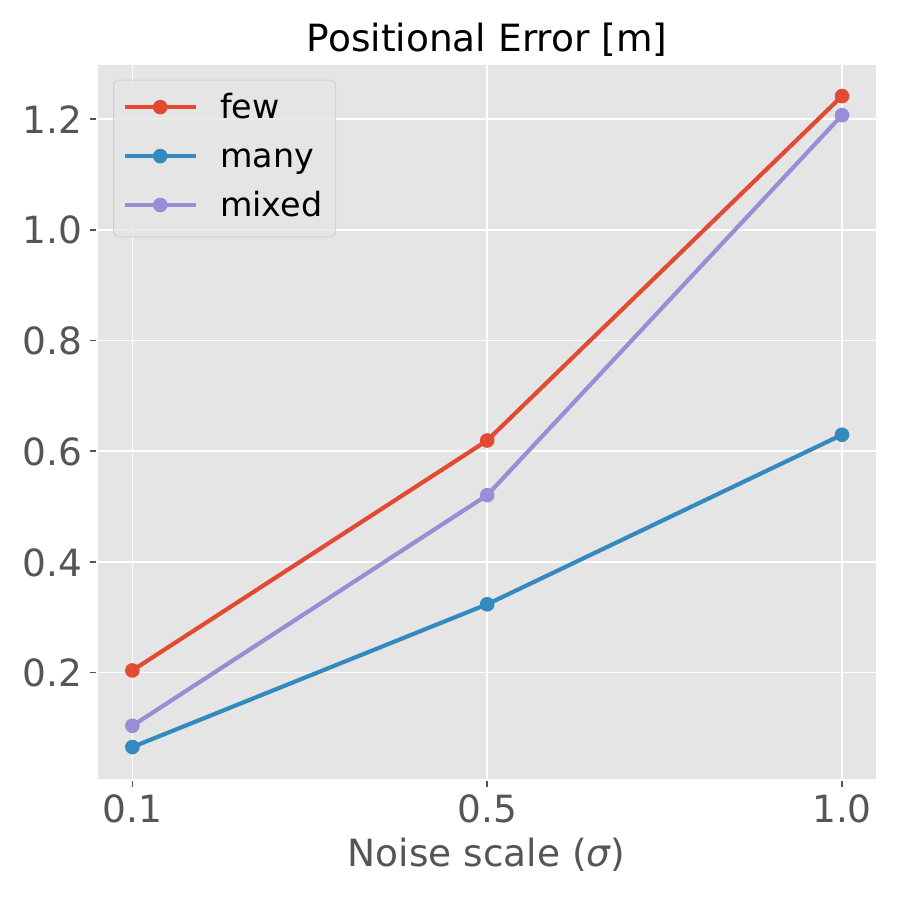}
\caption{Positional variability}
\end{subfigure}
\begin{subfigure}[b]{0.24\linewidth}
\includegraphics[width=\textwidth]{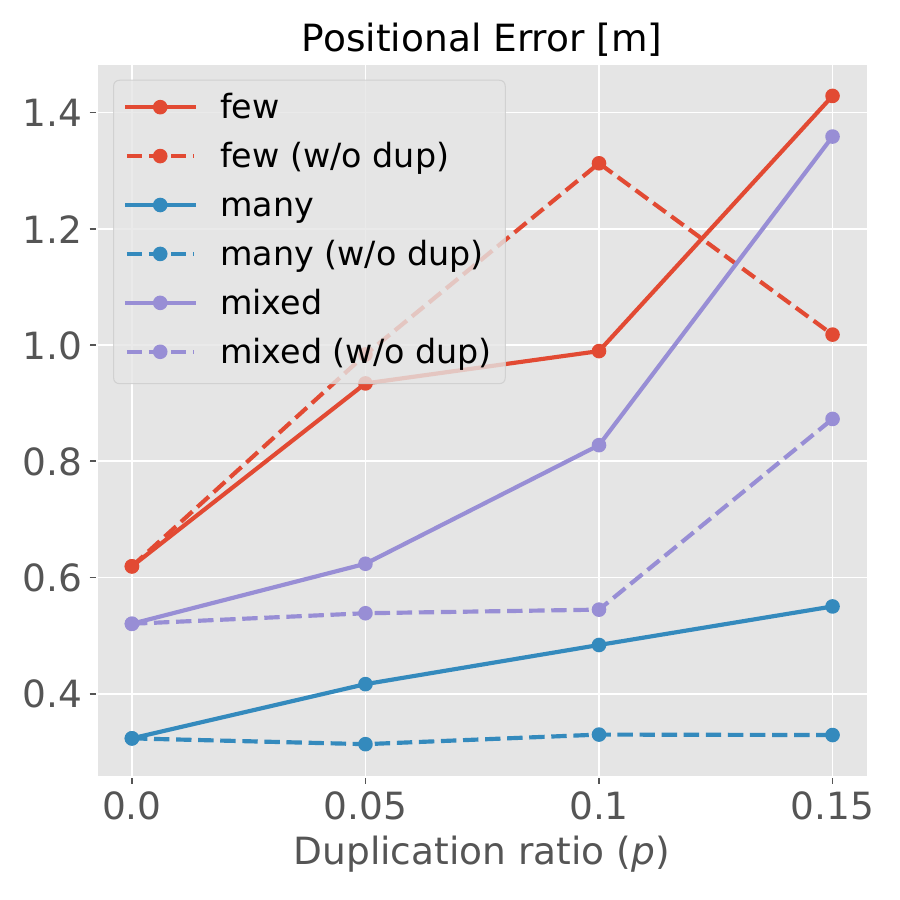}
\caption{Duplicated landmarks}
\end{subfigure}
\caption{Simulation experiments: (a) Effect of sampling conditions of landmarks for each data record, (b) effect of positional variability when arriving at landmarks, and (c) effect of the presence of duplicated landmarks as well as whether to ignore them in the aggregation.}
\Description{The figure consists of three sets of line charts, each analyzing different aspects of positional error and landmark identification in various scenarios. Chart set (a) titled "Landmark visits (few, many, mixed)" consists of two charts. The first chart shows positional error in meters as a function of the number of records, with three lines representing conditions: "few," "many," and "mixed" landmark visits. The second chart depicts the number of identified landmarks across the same conditions and records. The "few" condition shows higher positional error and fewer identified landmarks compared to "many" and "mixed." Chart set (b) titled "Positional variability" presents a single chart illustrating positional error in meters as a function of noise scale (σ). It compares the same conditions: "few," "many," and "mixed," with the "few" condition showing the highest error as noise increases. Chart set (c) titled "Duplicated landmarks" includes a chart showing positional error in meters relative to duplication ratio (ρ). It compares conditions with and without duplication (indicated by solid and dashed lines respectively). The conditions are "few," "many," and "mixed," with "few" showing higher error, especially when duplication is considered.}
\label{fig:sim_plot}
\end{figure}

\subsection{Effect of the Number of Landmarks per Recording}
First, we observed how the number $K$ of landmarks in each data record affects the positional error and landmark coverage, while changing the number of records to aggregate from $[3,6,9,12,15]$. The noise scale and landmark duplication ratio were fixed to $\sigma=0.5$ and $p=0$ (no duplication). Fig.~\ref{fig:sim_plot}a demonstrates that in the \textit{many} condition where each participant identified 12-15 landmarks, coverage reached approximately 25 out of 30 landmarks with just 3 records and approached almost 30 landmarks with 6 records. 
The positional error remained consistently low across all record counts, indicating that the substantial overlap in identified landmarks combined with our alignment algorithm effectively aligned the coordinate system for each recording. In contrast, the \textit{few} condition showed significantly slower growth in the number of identified landmarks, never reaching 30 landmarks. Interestingly, in the \textit{mixed} condition, even 10\% of records (\ie, only 1 or 2 records) with many landmarks greatly improved the positional error over the few condition. The result suggests that encouraging even a few users to visit around 40-50\% of the total number of landmarks per single recording helps stabilize localization performance. On the other hand, ensuring a high level of landmark coverage requires either most users to visit many landmarks or a significantly larger number of users to participate in the task.

\subsection{Effect of the Variability in Distances between Users and Landmarks} 
As we observed in the user study, participants described target landmarks from various positions, where the standard deviation of the distances from participants to landmarks was $\sigma\sim0.5$. In this experiment, we systematically varied the noise scale using three values: $\sigma \in [0.1, 0.5, 1.0]$ and observed how they affected the positional error after aggregation. Similar to our first simulation, we compared the few, many, and mixed sampling conditions while fixing the number of records to aggregate at 15. Fig.~\ref{fig:sim_plot}b demonstrates that the noise scale significantly affects performance across all sampling conditions. While omitted from the paper, the same trend was found for other choices of the number of records to aggregate. This result indicates that the proposed system is most suitable for environments where user movement is somewhat constrained, such as complex stores with narrower aisles rather than large shopping malls with many open spaces.

\subsection{Effect of Duplicated Landmarks}
Finally, we investigated the effect of duplicated landmarks by changing the parameter $p \in [0, 0.05, 0.1]$ where $p=0.1$ means 10\% of landmarks are assigned the same IDs as others. As depicted in Fig.~\ref{fig:sim_plot}c, processing all landmarks including those with duplicated IDs demonstrates degraded performance as the duplication ratio increases. This is expected because our aggregation algorithm would inevitably match landmarks from multiple records even when they have the same name but are physically located in different places. Nevertheless, we can mitigate this issue straightforwardly by ignoring landmarks with duplicated names during the alignment optimization step (\ie, removing landmarks with non-unique names when computing the transformations), as shown by the dotted lines in Fig.~\ref{fig:sim_plot}c. Such landmarks can be easily identified in practice, for example, when the same landmark appears multiple times in different locations within a single record or when we know actual sales plans to place the same product in multiple locations to compare sales performance.

\section{CONCLUSION}
We presented \codename, a map-as-a-byproduct system that creates semantic landmark maps from smartphone IMU data and user-provided texts during users' regular tasks. We confirmed the effectiveness of our system with retail restocking checklist tasks as well as office inspection tasks, demonstrating average positional errors of 0.8 m and 1.9 m, respectively, with nearly perfect landmark coverage. Our work contributes a novel, easy-to-deploy approach that enables indoor participatory sensing for collecting location-tagged semantic data in a semi-automated fashion by leveraging existing workflows and tools. This approach addresses key challenges in retail domains discovered in our interview study, particularly the need for up-to-date location-tagged semantic information without requiring dedicated hardware installations or significant additional operational burden.

\subsection*{Acknowledgments}
This work was supported by JSPS KAKENHI Grant Number 23K24831. The authors would like to thank Takeshi Ishita for conducting the formative interviews.
This research is supported by Singapore Ministry of Education (MOE) Academic Research Fund (AcRF) Tier 1 grant.

\bibliographystyle{ACM-Reference-Format}
\bibliography{sample-base}

\end{document}